# Systematic Review for Anomaly Network Intrusion Detection Systems: Detection Methods, Dataset, Validation Methodology, and Challenges

**Ziadoon K. Maseer[1], Robiah Yusof[2], Baidaa Al-Bander[3], Abdu Saif[4], Qusay Kanaan Kadhim[5]**

[1]Faculty of Computer Technology Engineering, Bilad Al Rafidain University College (ziadoonkamil@gmail.com)
[2]Faculty of Information and Communication Technology, Universiti Teknikal Malaysia Melaka, Malaysia (robiah@utem.edu.my)
[3]School of Computing, Keele University, UK (b.al-bander@keele.ac.uk)
[4]Faculty of Engineering, Taiz University, Taiz, Yemen (abdu.saif@taiz.edu.ye)
[5]Department of Computer Science, College of Science, University of Diyala, Baqubah 32001, Diyala, Iraq (qusaykn@gmail.com)

Corresponding author: Ziadoon K. Maseer[1] ( ziadoonkamil@gmail.com)

**ABSTRACT:** Intrusion detection systems (IDSs) built on artificial intelligence (AI) are presented as latent mechanisms for actively detecting fresh attacks over a complex network. Although review papers are used systematic review or simple methods to analyse and criticize the anomaly NIDS works, the current review uses a traditional way as a quantitative description to find current gaps by synthesizing and summarizing the data comparison without considering algorithms performance. This paper presents a systematic and meta-analysis study of AI for network intrusion detection systems (NIDS) focusing on deep learning (DL), and machine learning (ML) approaches in network security. Deep learning algorithms are explicated their structure and data intrusion network is justified based on an infrastructure of networks and attack types. By conducting a meta-analysis and debating the validation of the DL and ML approach by effectiveness, used dataset, detected attacks, classification task, and time complexity, we offer a thorough benchmarking assessment of the current NIDS-based publications-based systematic approach. The proposed method is considered reviewing works for the anomaly-based network intrusion detection system (anomaly-NIDS) models. Furthermore, the effectiveness of proposed algorithms and selected datasets are discussed for the recent direction and improvements of ML and DL to the NIDS. The future trends for improving an anomaly-IDS for continuing detection in the evolution of cyberattacks are highlighted in several research studies.

## I. INTRODUCTION

According to Cybersecurity Ventures, the overall volume of data gathered in the cloud comprises public clouds operated by industrial and social media companies (for instance, Twitter, Microsoft, Google, Facebook, Apple, etc.) [1],[2]. Consumers and corporations can utilize government-owned cloud services. Mid-to-large businesses own cloud storage providers and private clouds [3] . By 2025, data will have achieved 100 zettabytes, and worldwide data storage will have surpassed 200 zettabytes [4] .Data stored on public and private network infrastructure data centers and private computing devices, for instance, Internet-of-Things (IoT) devices, smartphones, and PCs, are included in the cloud. Given the interchange of vast volumes of sensitive information through resource-constrained devices and across the untrusted Internet utilizing communication protocols, unknown IoT devices and heterogeneous technologies, this fast expansion creates significant security issues. Robust security controls and resilience analysis should be performed in the early stages before installation to ensure secure and sustainable cyberspace. To preserve these technologies progressing, the implemented security controls are accountable for deterring, identifying, and resolving attacks [5]. With an increasing number of devices numerous companies will be limited resources on network defense to preserve their systems from intrusion, and they practice conventional defense methodologies and instruments such as antispam, antivirus, firewalls, etc [6] .However, these methodologies and devices are exposed to novel attacks [7]. Network intrusion detection systems (NIDS) are thus critical tools for recognizing intrusions, tracking malicious actions, and monitoring network traffic. NIDS implements robust protection systems facing numerous threats. Furthermore, IDS techniques are categorized into signature and anomaly approaches depending on their work processes [8]. The anomaly IDS identifies abnormal activities from regular routines to identify the connection lines as usual or intrusive behavior. At the same time, the approach will be affected to protect the network system against unknown attacks [9]. Hence, signature IDS models rely on predetermined attack criteria and a wide range of network traffic, including platforms, social applications, and IoT media [10], [11]. Meanwhile, it will be unrealistic to rely solely on recognizing predefined patterns of attacks to detect intrusions. Therefore, anomaly-based IDS models are essentially employed to know the unknown attacks for predefined patterns in the case of the absence of IoT networks [12]. Several ML and DL approaches were proposed in the latest decade to enhance the NIDS's effectiveness in exposing ill-disposed crimes. Furthermore, massive improvements have been done to develop current industries such as data center , medical care IT [13], and 5G services [14]. The emergence of new threats could lead to significant difficulties in discovering new attacks within modern network by NIDS models.

Based on Table 1, there are still weak points in prior articles review to investigate the gaps and latest trends for anomaly NIDS. A Few articles offered to review only trends future work of anomaly detection without a systematic review [8, 9, 10, and 13], and other authors had submitted to discuss the performance of anomaly NIDS by using a systematic approach based on deep learning methods [14 -17]. Where quantitative research is a process of together numeric data from testing results, ML/DL-NIDS models had tested using the quantitative method or approach due to data traffic is a numeric value. A quantitative method is making a clear investigation to find out gaps and future trends by meta-analysis with anomaly NIDS models [10], [49]. Thus, there remains an enormous opportunity to review articles regarding the effectiveness and detected attack types to give a clear picture of recent trends and gaps for NIDS using meta-analysis or quantitative methods.

We propose a comprehensive systematic meta-analysis for reviewing the state-of-art anomaly-based NIDSs, to an assisted stable and reliable network in the security domain. The critical aspect of our approach is to select current issues in the meta-analysis for the studies from 2017 to 2022 that should be the future trends in the security domain of NIDs. Therefore, the contributions are summarized as follows:

**TABLE 1: COMPARISON OF REVIEW ARTICLES: (✓: YES, ×: NO)**

| Review Article | Year | AIDS | ML | DL | Systematic Study | Meta-analysis | Current Issues | Future Trends |
|---|---|---|---|---|---|---|---|---|
| [15] | 2017 | IDS | × | ✓ | × | × | × | ✓ |
| [16] | 2017 | IDS | × | × | × | × | × | ✓ |
| [17] | 2020 | IDS | × | × | × | × | × | ✓ |
| [18] | 2020 | Anomaly-IDS | × | ✓ | × | ✓ | × | ✓ |
| [19] | 2019 | Anomaly-IDS | ✓ | ✓ | × | × | × | × |
| [20] | 2019 | Anomaly-NIDS | ✓ | × | × | × | ✓ | × |
| [21] | 2020 | Anomaly-NIDS | ✓ | ✓ | ✓ | × | × | ✓ |
| [22] | 2021 | IDS | × | × | ✓ | × | × | × |
| [23] | 2021 | IDS | × | × | × | × | × | ✓ |
| [24] | 2021 | IDS | × | × | × | × | ✓ | ✓ |
| [25] | 2021 | IDS | ✓ | ✓ | × | × | × | ✓ |
| [26] | 2022 | Anomaly-NIDS | ✓ | × | ✓ | × | × | × |
| [27] | 2022 | Anomaly-NIDS | × | × | ✓ | × | × | × |
| **This Article** | 2023 | Anomaly-NIDS | ✓ | ✓ | ✓ | ✓ | ✓ | ✓ |

**Research Domain:** According to NIDSs released over the previous four years, we collect data and present studies that have obtained by systematic research to DL and ML from 2017 to 2022. Furthermore, the structure of DL and ML methods are explained and, also available datasets are identified and explained based on the development history of data traffic for training anomaly NIDSs in the security domain.

**Quantitative analysis:** We further investigate the performance of ML NIDS based on the proposed methodology, meta-analysis, datasets, detected attacks, and efficiency, creating quantitative research. We provide a more accurate status of using ML/DL approaches for NIDS. We then highlight various current issues and future trends in ML/DL NIDSs in this area of network security.

The rest of this paper is organized as follows: the research approach applied in this area is discussed in Section 2. Section 3 contains reviews for supplementary investigation of classification strategies in anomaly-NIDS detection methods. The network's public dataset, which was intended to train and test the anomaly-NIDS model, is presented in Section 4. Section 5 presents the different factors related to network security, summarizing information from the research's growth of NIDS performance, such as effectiveness, datasets, task classification, hardware implementation, and execution time, as shown in Appendix A. Besides, it provides in-depth knowledge and statistics for current algorithms and datasets proposed for the NIDS using ML/DL. Section 6 presents the current challenges and highlights future trends. Table 2 lists of abbreviations used throughout this paper:

## II. ARTICLE SELECTION METHODOLOGY

This section presents a methodology to implement a systematic literature review (SLR) based on a set of steps to identify, examine, and summarize valuable knowledge from state-of-the-art related to genuine research topics. For the published journal papers from 2017 to 2022, the various DL- and ML-based NIDSs are evaluated and studied. A systematic literature review is developed and organized for performing data extraction, Assessing, synthesizing, and analyzing results into four phases, illustrated in Figure 1.

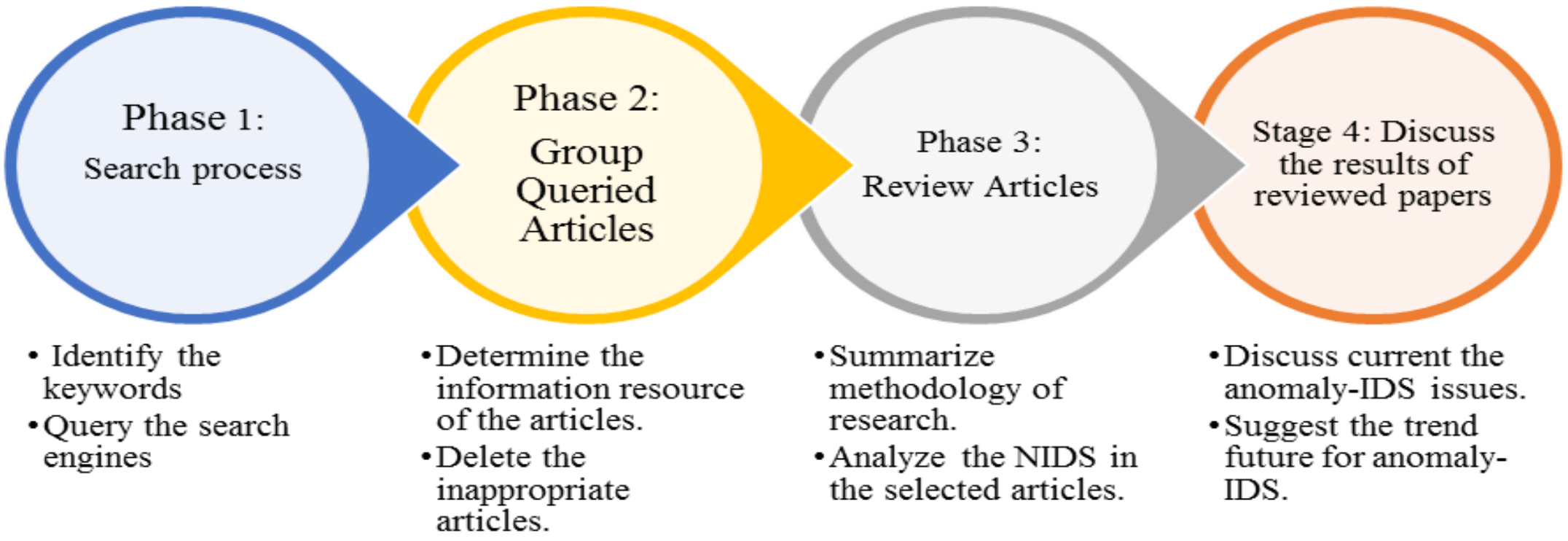

**FIGURE 1: Systematic Review Process**

**TABLE 2: LIST OF ABBREVIATIONS**

| Abbreviation | Description |
|---|---|
| AI | Artificial Intelligence |
| DL | Deep Learning |
| ML | Machine Learning |
| ANIDS | Anomaly Network Intrusion Detection |
| SL | supervised learning |
| ANNs | Artificial Neural Networks |
| BP | backpropagation |
| FFNN | forward neural network |
| DT | Decision Tree |
| K-NN | K-Nearest-Neighbor |
| NB | Naive Bayes |
| RF | Random Forest |
| SVM | Support Vector Machine |
| EM | Expectation-Maximization |
| SOM | Self-Organizing Map |
| DNN | Deep Neural Network |
| RBM | Restricted Boltzmann Machine |
| DBN | Deep Belief Network |
| AE | Auto Encoder |
| CNN | Convolutional Neural Network |
| RNN | Recurrent neural networks |

**Phase 1:** Considering its capabilities to investigate all identified databases, Google Scholar was selected as the default search engine. We determined a search query using words recognized for research. The database will be searched using the following keywords related to IDS research: “anomaly” AND “intrusion detection system” from 2017 to the first quarter of 2022. These keywords are the most proper guideline in this domain.

**Phase 2:** With the initial results of the search query in google scholar based on title research, it was 17,500 papers. We identified journal articles published between 2017 and 2022. There are random works related to anomaly-NIDS models from different resources. We identify good resources in computer science and information technology: *Springer, Elsevier, IEEE and Scholar*. The references are examined and commented on by scientists with details related with anomaly IDS in measurements, datasets, and methodology. The essential papers are selected to review and analyze based on relayed information in this study, as depicted in Table 2.

**TABLE 2: PARAMETERS SELECTION FOR NIDS ARTICLES**

| Data retrieved | Description |
|---|---|
| Title | Title of the main study |
| Journal | Name of the Journal publishing the article |
| Application area | The area within which the case has been studied |
| Approach | The proposed approach for IDS |
| Method | The algorithm implemented to conduct the study |

To reduce the selected papers from the initial set and produce this systematic review paper, we defined inclusion and exclusion criteria. Inclusion criteria included papers focusing specifically on anomaly-based NIDS, papers published within a certain time frame (2017- 2022) to ensure relevance, papers written in English language for ease of understanding, papers reporting original research studies, including empirical studies, experiments, simulations, or case studies, papers presenting novel techniques, algorithms, or approaches related to anomaly-based NIDS, papers that evaluate the performance or effectiveness of anomaly-based NIDS using appropriate evaluation metrics, papers providing detailed methodologies and technical descriptions of the anomaly-based NIDS, papers discussing real-world deployment, implementation, or practical considerations of anomaly-based NIDS, papers that provide insights into the challenges, limitations, or future directions of anomaly-based NIDS, and papers from reputable journals, conferences, or academic sources to ensure credibility. Whereas exclusion criteria included papers focusing on other types of intrusion detection systems rather than anomaly-based or behavior-based NIDS, papers not directly related to the research question or objective of the systematic review, papers without access to the full text (e.g., abstract-only papers, inaccessible conference proceedings), papers that primarily discuss general network security topics without specific emphasis on anomaly-based NIDS, papers that are duplicates or multiple publications of the same study, papers lacking sufficient details or methodology sections to assess the quality and rigor of the study, papers that are primarily theoretical discussions or opinion-based without empirical evidence or evaluations, papers not published in recognized academic or peer-reviewed sources.

We conducted an initial screening by screening the titles and abstracts of the papers to identify those that potentially meet the inclusion criteria. This initial screening helped to eliminate papers that were obviously irrelevant to the research question. After the initial screening, we carefully evaluated the full texts of the remaining papers. We thoroughly read each paper and compared it against the inclusion and exclusion criteria. Papers that clearly do not meet the criteria were excluded at this stage.

**Phase 3:** As we reviewed the full-text papers, we extracted relevant data such as the research objectives, methodology, findings, and conclusions. We created a structured form to record this information. Spreadsheet tools (Microsoft Excel) were used for data extraction during a systematic review. We created customized templates with relevant fields to record information from selected studies. Based on the extracted data, we categorized the papers into different groups or themes (based on the NIDS techniques employed, evaluation methods used, datasets utilized, etc.). Each appropriate research is summarized and investigated for the suggested DL/ML-based NIDS methodology. The parameters are centered on the most often utilized datasets, testing metrics, and whether the categorization might be a multi- or binary task to evaluate and analyze the works. The essential criterion concerns the classification task for NIDS because most datasets are imbalanced classes, and IDS software is designed to detect multi-attacks. The software will be more general than designed software for binary attacks. Therefore, we established the accuracy of the suggested attack detection algorithms. Moreover, the time complexity is an essential factor in measuring the time of algorithms required to complete the task of the NIDS to avoid overhead and packet loss [28]. With those criteria, we provided a significant study of current NIDSs based on ML/DL.

**Phase 4**: Lastly, we utilized Appendix A summarized information findings in reviewed articles for forming the future trends and challenges of this research for AI-based NIDSs.

## III. BACKGROUND NIDS

Anomalies refer to patterns, events, or observations that deviate significantly from the expected or normal behavior in a given context. In the context of anomaly detection, anomalies are considered as instances that differ significantly from the majority of the data or exhibit behaviors that are unusual or suspicious. Anomaly detection involves identifying and flagging such abnormal instances or patterns within a dataset or system. Anomaly detection, also known as outlier detection, is the process of identifying these unusual or unexpected patterns within a dataset or system. The goal is to distinguish anomalies from the majority of normal instances and potentially identify potential threats, fraud, errors, or unusual behavior that might require further investigation or action. The use cases for anomaly detection are diverse and span various domains including network intrusion detection [29] ,fraud detection [30] , manufacturing and quality control [31] , cybersecurity [32], health monitoring [33] , and IoT device monitoring [34].

Network Intrusion Detection Systems (NIDS) are a crucial component of network security infrastructure. They are designed to identify and respond to malicious activities or unauthorized access attempts within computer networks. NIDS play a vital role in detecting and preventing network-based attacks, providing a proactive defense mechanism against cyber threats. The background of Network Intrusion Detection Systems can be traced back to the increasing prevalence and sophistication of network attacks. As computer networks became more interconnected and vital for organizations, attackers started exploiting vulnerabilities to gain unauthorized access, steal sensitive information, disrupt services, or execute malicious activities. The need for effective intrusion detection mechanisms led to the development of NIDS. NIDS are software or hardware systems that monitor network traffic in real-time, analyzing it for signs of suspicious or malicious behavior. They work by inspecting network packets, network protocols, and traffic patterns to identify potential threats and attacks.

There are two primary approaches to NIDS: signature-based and anomaly-based detection. Signature-based NIDS rely on a database of known attack patterns or signatures. They compare the network traffic against these signatures to identify matches and raise alarms when a known attack is detected. Signature-based NIDS are effective in detecting known attacks but may struggle with new or evolving attack techniques that do not match existing signatures. On the other hand, anomaly-based NIDS focus on establishing a baseline of normal network behavior and identifying deviations from this baseline. They use statistical analysis, machine learning, or rule-based approaches to detect abnormal patterns or behaviors that may indicate an ongoing attack. Anomaly-based NIDS can detect novel attacks but may have a higher false positive rate due to the inherent challenges in distinguishing between legitimate and malicious anomalies [31, 32].

NIDS are typically deployed at strategic points within a network, such as at network gateways, routers, or switches. They monitor network traffic in real time and generate alerts or take automated actions when suspicious activities are detected. NIDS can detect various types of network attacks, including network scanning, port scanning, denial-of-service (DoS) attacks, malware propagation, and unauthorized access attempts. Over time, NIDS have evolved to incorporate advanced techniques, such as deep packet inspection, behavioral analysis, and threat intelligence integration. They have become an integral part of network security architectures, working alongside other security components like firewalls, intrusion prevention systems (IPS), and security information and event management (SIEM) systems. The continuous development and improvement of NIDS are driven by the ever-evolving threat landscape and the need for robust network security. As attackers employ sophisticated techniques, NIDS must adapt and enhance their detection capabilities to ensure the early identification and mitigation of network-based threats [37].

## IV. DETECTION METHODS OF NIDSs

In general, learning to train machine and deep learning ML/DL algorithms can be either supervised or unsupervised. Algorithms belonging to supervised learning are those trained by classifying cases based on their data labels and then continuing to learn until reaching optimal or maximum value criteria with minimum loss. The supervised learning algorithms in conventional ML are the SVM, RF, NB, k-NN, DT, and ANN. Data instances that are not labeled can be found in unsupervised learning, in which clustering dominates the learning approach. SOM, EM, and K-means are unsupervised learning algorithms. Another anomaly-IDS technique is the DL approach, which has robust detection compared to machine learning by extracting features by defining attacks such as DNN, DBN, CNN, and RNN. These types of algorithms are portrayed in Figure 2.

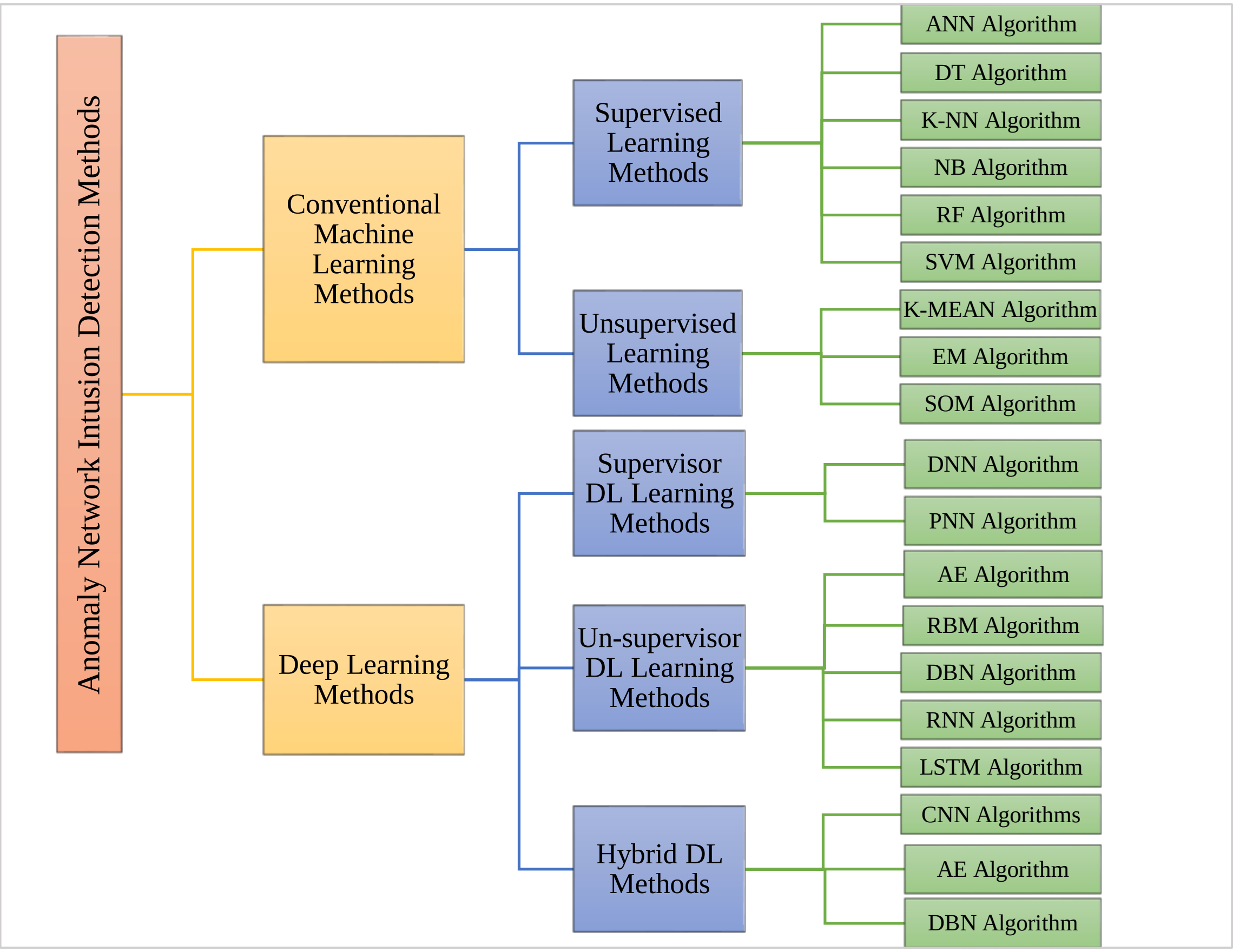

**FIGURE 2: Anomaly NIDS Methods**

### 1. SUPERVISED LEARNING METHODS

ML learning, a function task that translates input to an output premised on sample input-output pairs, is known as supervised learning (SL). In the traditional ML-NIDS, six supervised ML algorithms are evaluated. The following subsections go through the underlying notions of these algorithms in more detail:

### A. Artificial Neural Networks (ANNs)

ANNs are computing systems that model nonlinear problems and foresee output values based on their training values. The ANN comprises three aspects: weights, edges, and nodes or neurons to learn things and make decisions in a human-like manner. An ANN has three layers for constructing the ANN architecture: the output, hidden, and input layers, and each layer comprises a set of nodes. Each neuron's directed connection is affiliated with its weight and connects to other nodes in the next layer via an edge. The output layer represents a machine's or graph's outputs, and the output neurons are equal to labels of data training. The activation functions and weights construct the hidden layers to help discover the network security domain's underlying or structural data features [38]. Figure 3 depicts a general ANN structure (I-H-O) for conventional learning. (I) denotes the input layer nodes, H denotes the hidden layer nodes, and O denotes the output layer nodes. According to [39], an ANN algorithm for detecting unknown attacks used KDD CUP 99 datasets to train a multi-layer forward neural network (FFNN) and the mean loss function to reduce the errors. Another approach of an ANN is backpropagation (BP), which uses a backward mechanism from the last layer to update the hidden layers' biases and weights. The significant research works proposed to solve the low detection predictor found in an anomaly-NIDS using a neural network algorithm because of the over-or underfitting phenomenon [40]. Feature selection was presented to remove irrelevant features for enhancing the training of a neural network algorithm using the NLS-KDD dataset [41]. *K*-fold cross-validation is another solution to handle the overfitting issue by counting the average accuracy within training. Besides this, the meta-heuristic or heuristic algorithms effectively reduce the overfitting issue by selecting optimal parameters such as weight, bias, and the number of hidden neurons [42].

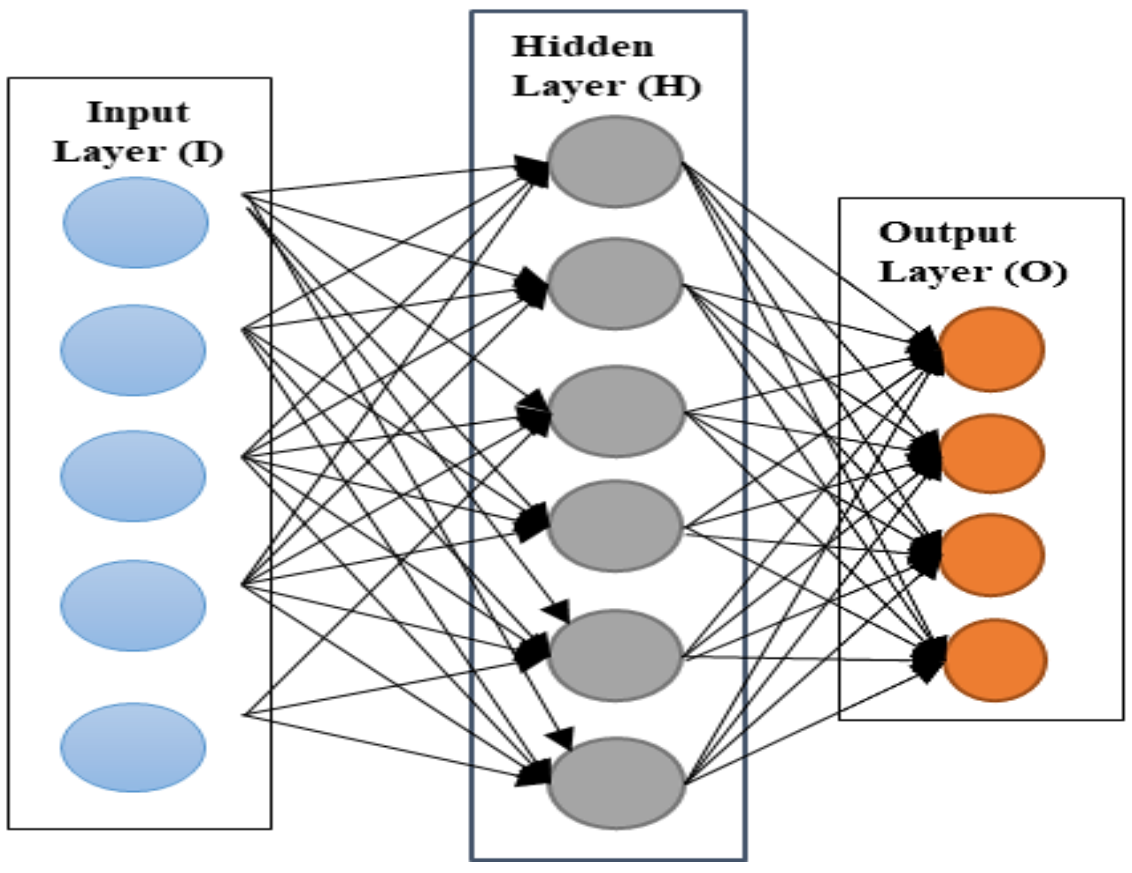

**FIGURE 3: ANN Architecture**

### B. Decision Tree (DT)

A decision support tool, or DT, is a collection of supervised learning algorithms often deployed to tackle machine learning categorization problems. A tree-like model of decision types relates to a circumstance in which the target variable accepts discrete values as input, known as classification trees. DTs comprise leaves or nodes, branches, and one root. While nodes represent the classes of the dataset, branches are the subset of features used to indicate the class labels. A DT may learn for both continuous and discrete data. Given an essential splitter in input determinants, the DT algorithm divides the attributes into two or more analogous sets. The overfitting problem is encountered by a DT, which is overcome via sampling methods, boosting, and bagging [43]. A common structural example of a DT is indicated in Figure 4 [44]:

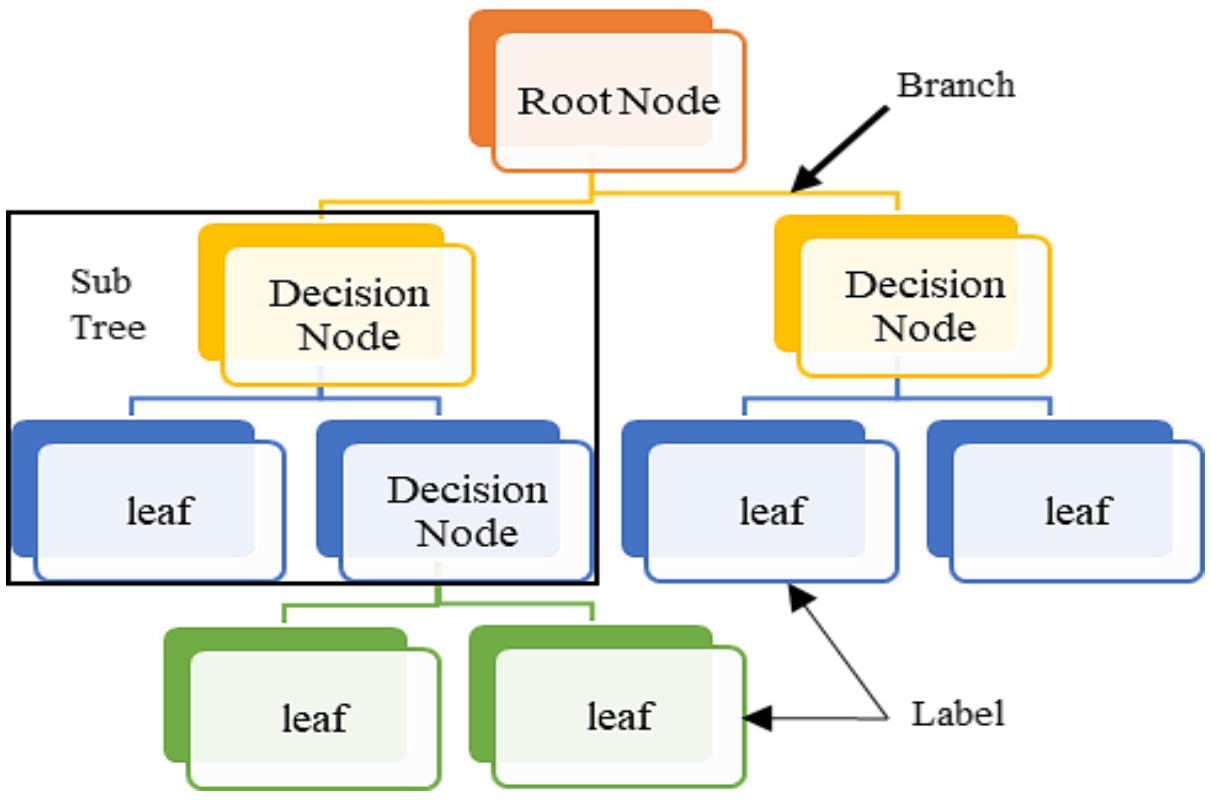

**FIGURE 4: DT Architecture**

### C. K-Nearest-Neighbor (K-NN)

The KNN algorithm is one of the most basic ML algorithms that may be utilized to tackle both regression and classification issues. It is based on the supervised learning technique. According to [45], the model can be a regression or classification method. The primary work of the k-NN algorithm uses the distance function to compute the similarity behavior of points or differences between a pair of points, denoted as D(a, b) in Equation 1 [46]:

$$D(a, b)=\sqrt{\sum_{i=1}^{r}(ai-bi)^2} \quad (1)$$

In which $a_i$ denotes the $i^{th}$-featured element of the instance a, $b_i$ is the $i^{th}$-featured element of the instance b, while r is the dataset that features the entire quantity. It represents a non-frontier approach with no intuition to publish the underlying data. This algorithm is a simple training of ML models based on the dataset because it requires a small dataset to find the distance between instances or instances and labels. At the same time, a more significant part of the dataset is used for testing this type of model. This is useful as the bulk of the actual dataset is effortlessly derived compared to what is done in a lazy algorithm as the parameter model. It assists the training stage, yet the testing stage is high in speed and memory and less generalized cases for detecting attacks.

### D. Naive Bayes (NB)

The NB methods belong to a simple group of probabilistic algorithms established by the Bayes theorem. It addresses naive conjectures of feature independence and dependent features with dataset labels [20]. It is easily trained through the utilization of a supervised learning structure representing the Bayes equation:

$$P(A|B) = \frac{P(B|A)\,P(A)}{p(B)}. \qquad (2)$$

represents the labels or dependent events, and B represents the values of features. The previous label probability denotes P(A), and the features prior probability denotes P(B), both of which must not be zero. Provided that hypothesis A is true, P(A|B) is the posterior probability of B, and P(B|A) is the probability of the features.

### E. Random Forest (RF)

An RF is a simple method to build a model using the bagging technique to solve a DT experiencing overfitting and detection issues. An RF addresses this matter by enabling the middle of deep decision trees [47]. An RF is an approach to solving regression and classification problems utilizing an ensemble learning approach. Its functions create multiple DTs within the training stage using split data as the bootstrap part. The output considers each feature of the dataset as the classes' mode of a particular DT during the execution of a classification function. Figure 5 shows the RF architecture based on forest trees.

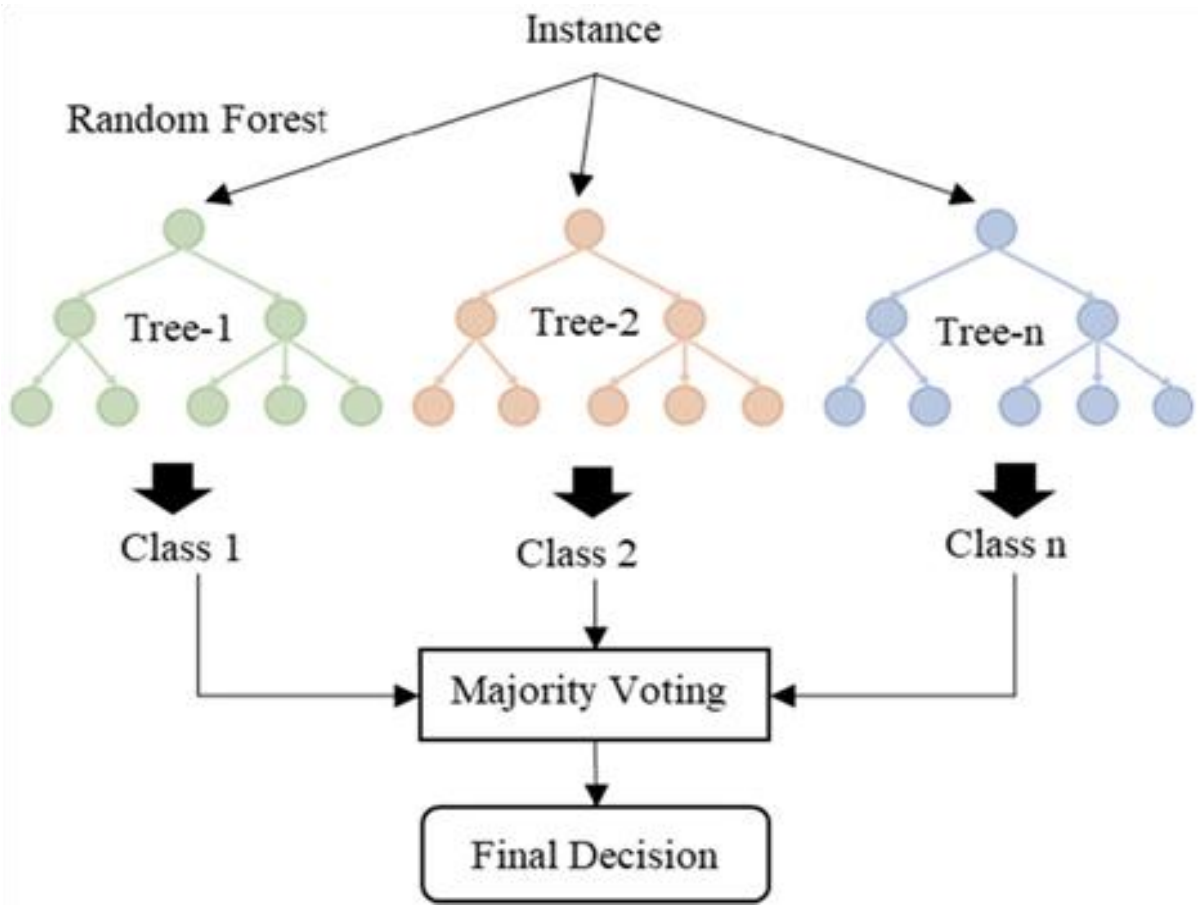

**FIGURE 5: RF Architecture**

### F. Support Vector Machine (SVM)

It belongs to supervised and linear learning that uses a plane that classifies the instances into varying categories. Numerous planes can separate the training instance sets. However, the optimal plan finds support vectors with the highest distance or margins from any class's nearest instances and plane. A hyperplane can be used as a prediction, as presented in Equation 3.

$$g(x) = \begin{cases} +1 \text{ if } w.x + b \geq 0 \\ -1 \text{ if } w.x + b \leq 0 \end{cases}, \qquad (3)$$

g(x) is the predicted class if g(x) is more significant than zero as the normal class, else g(x) is less than zero as the abnormal class [48]. Figure 6 shows an optimal hybrid plane to classify data based on maximized margins for points or support vectors. The disadvantages of linear SVM are that it cannot classify nonlinear data, has a low accuracy with noise, and has an overlapping issue. Furthermore, training SVM with a large dataset takes a long time. The genetic algorithm was proposed as a feature selection to select significant features to overcome the disadvantage of the SVM classifier [49]. Other research proposed SVM and K-NN as a preprocessing stage to classify data to improve training the weighted majority algorithm for the generating model [50].

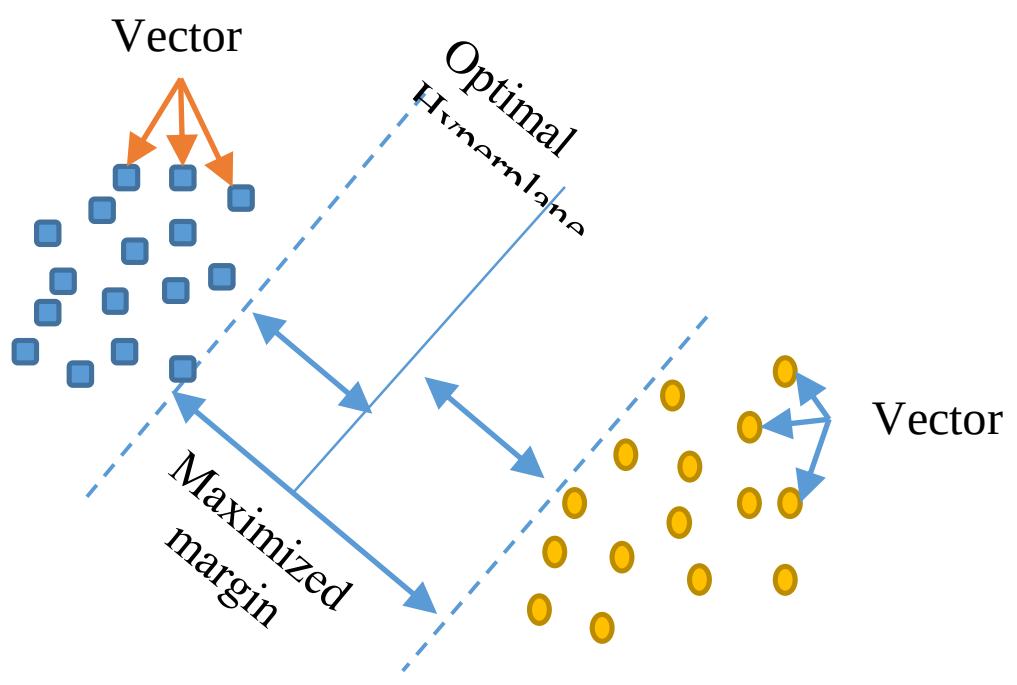

**FIGURE 6: SVM Architecture**

## 2. UNSUPERVISED LEARNING ALGORITHMS

Unsupervised learning is a sort of ML that utilizes an unlabeled dataset and operates on it without being supervised. There are three selected algorithms as basic concepts of unsupervised machine learning, and they are further discussed in the following subsections:

### A. K-Means Clustering

K-means clustering represents an unsupervised learning method for classifying identities into a fixed number (k) of clusters utilizing an unlabeled dataset. It is among the common simple unsupervised learning approaches according to the outlook distance of points. It divides the n samples into k groups, and each instance is linked to the cluster that occupies the closest mean with similar behavior. The main disadvantage of this algorithm is that it needs the pre-specification of the number of clusters k. Provided that a set of instances ($p_1$, $p_2$,…, $p_n$), in which each instance represents a d-dimensional real vector, k signifies clustering that targets to partition p instances into $(k \leq p)$ sets $Z = \{Z_1, Z_2, \ldots, Z_k\}$ to minimize the variance. Then, k-means is expressed as equation 4 [51][52].

$$a_z \min \sum_{i=1}^{k} \sum_{p \in Zi}^{max} ||p - m_i|| \, a_z \min \sum_{i=0}^{k} |Z_i| \, \mathrm{Var} Z_{i.} \qquad (4)$$

A denotes an argument, while µi denotes the mean of points in set $Z_i$.

### B. Expectation-Maximization (EM)

Expectation-maximization algorithm performs maximum likelihood estimation with incomplete data, missing and unobserved (hidden) latent variables for expectation-maximization such as k-means. [53]. Thus, the EM algorithm computes cluster membership probabilities based on one or more distributions. Given the final clusters, it aims to maximize the data's overall probability.

### C. Self-Organizing Map (SOM)

A SOM is an unsupervised neural network that employs a map to reflect the input distribution's dimensionality. It is predicated on a neural network model known as unsupervised learning. A SOM can cluster data without prior knowledge of input data class groupings [54]. Its research adds a topology mapping from high-dimensional data created by mapping neurons known as units. The mapping preserves the distance between samples, for instance, the most likely closest locations, which reflect close map units. The SOM can define the data that have occurred before. Figure 7 demonstrates the SOM structure.

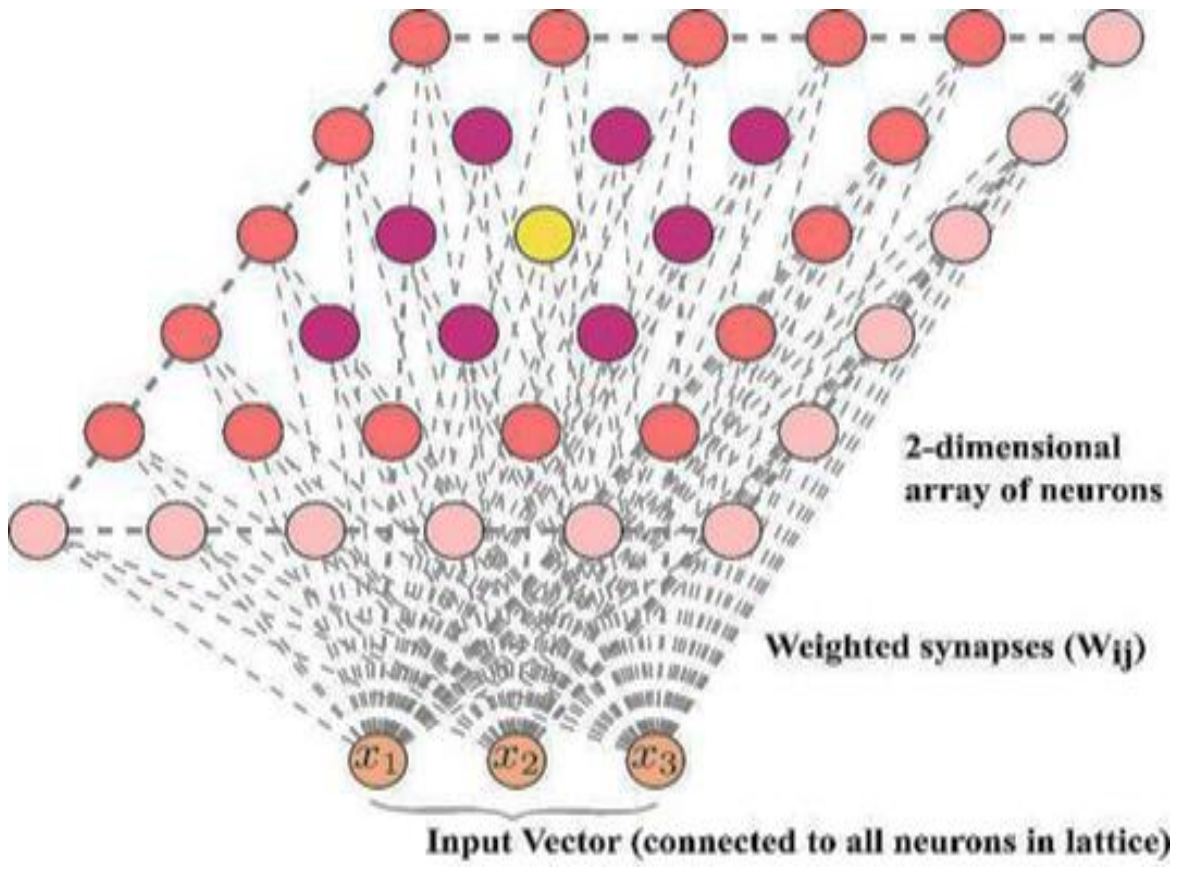

**FIGURE 7: SOM Architecture**

## 3. DEEP LEARNING APPROACH

DL is an ML technique that teaches computers to do what comes naturally to humans. There are seven algorithms involved in DL approaches. The fundamental theories of these algorithms are further discussed in the following subsections:

### A. Deep Neural Network (DNN)

An ANN with several layers between the output and input layers is known as a DNN. Deep learning refers to neural networks having more than three layers and more than one hidden layer. Nowadays, the number of network layers utilized in deep learning ranges from five to over a thousand. As illustrated in Figure 8, DNNs can learn high-level features with greater complexity and abstraction than shallower neural networks.

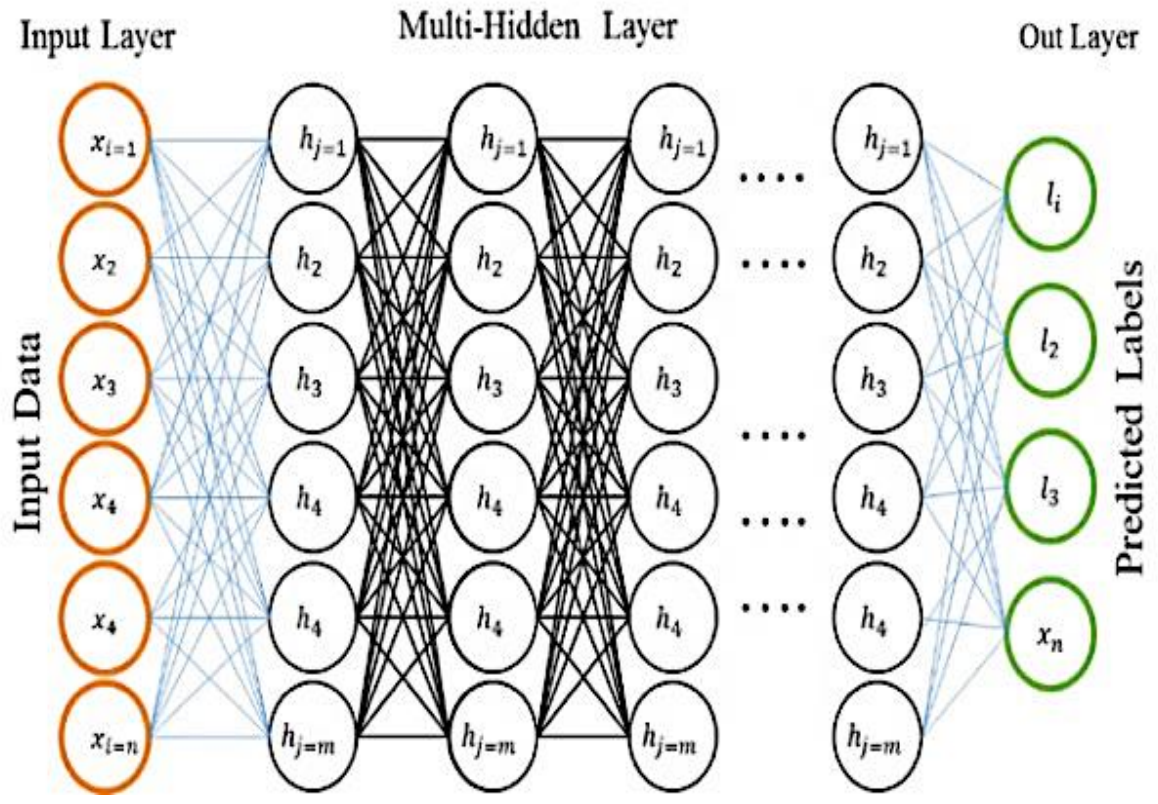

**FIGURE 8: DNN Architecture**

DNNs are used to process ADFA-WD data to display this point [55]. The data application samples are supplied into the first layer of a DNN in this context. The layer's outputs can indicate various low-level label properties, for instance, attacks and normal. Thus, these traits are integrated to determine the likelihood of higher-level features present at succeeding layers. Moreover, considering all this data, the network probability (comprised of these high-level attributes) is a specific scene or object in the last stage. DNNs can offer improved detection rate performance due to this deep feature hierarchy.

### A. Restricted Boltzmann Machine (RBM)

An RBM is a two-layer neural network. There are input or visible and hidden layers. Nodes in a visible layer are connected to nodes in a hidden layer, as shown in Figure 9

The nodes inside the hidden layer are also backpropagated to the visible layer in a traditional Boltzmann machine (BM). To reduce computational complexity, the neurons of a layer do not connect in an RBM. RBMs use a stochastic technique to calculate the probability distribution of the training dataset throughout the training phase. Once the training commences, each node is given random values. There are also prejudices in the visible and hidden layers of the model. The data is sent to the visible neurons or units that aid in rebuilding the input as it is being transferred to develop the hidden layer's activations. The model's reconstructed input does not match the data input. Generative models are another term for them.

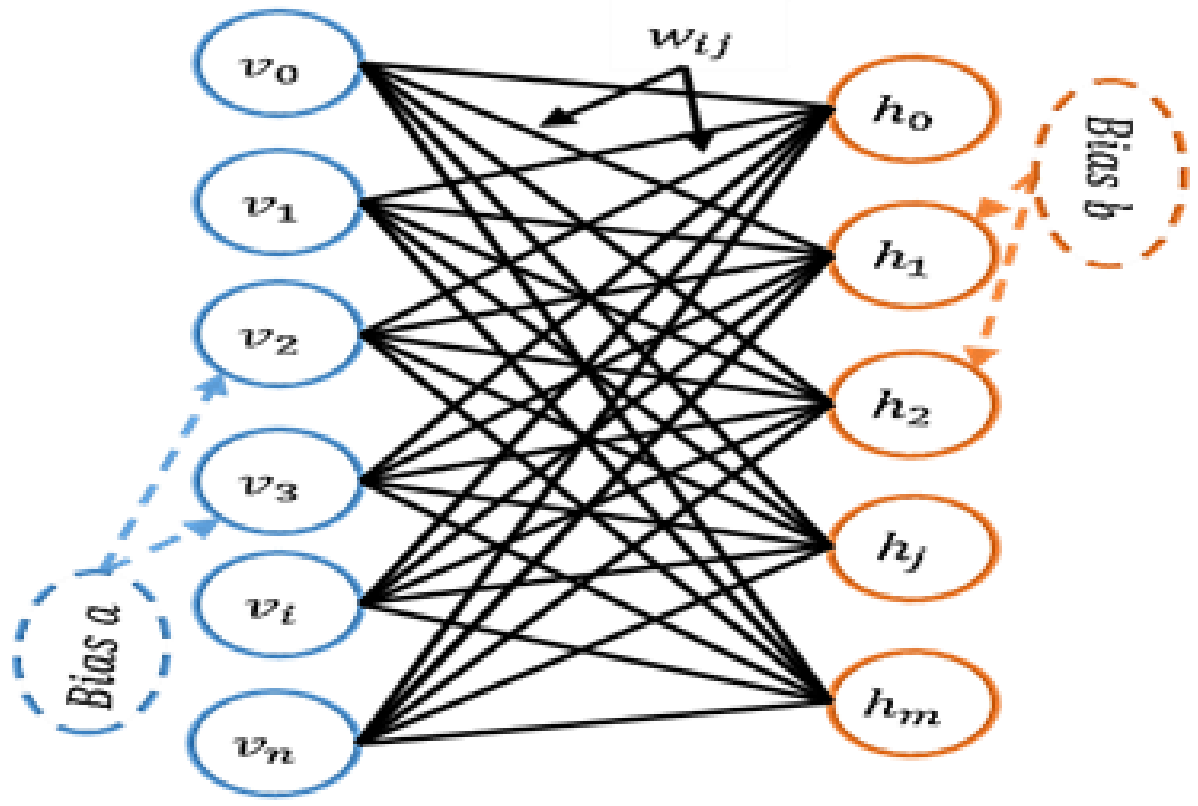

**FIGURE 9: RBM Architecture**

*B. Deep Belief Network (DBN)*

A DBN algorithm is a layer-wise approach for discovering data characteristics from unlabeled features unsupervised that are heavy or greedy [56]. The DBN is made of RBMs; each RBM's output becomes the input to the next RBM. The pre-training of the DBN shows how to enhance the ANN by training each layer separately to solve shallow learning. The DBN layers are separately trained until they reach optimal parameters. As illustrated in Figure 10, a DBN [57] may be utilized to categorize detected features in supervised learning by combining two backpropagated layers to construct the classifier and the final RBMs to classify detected features.

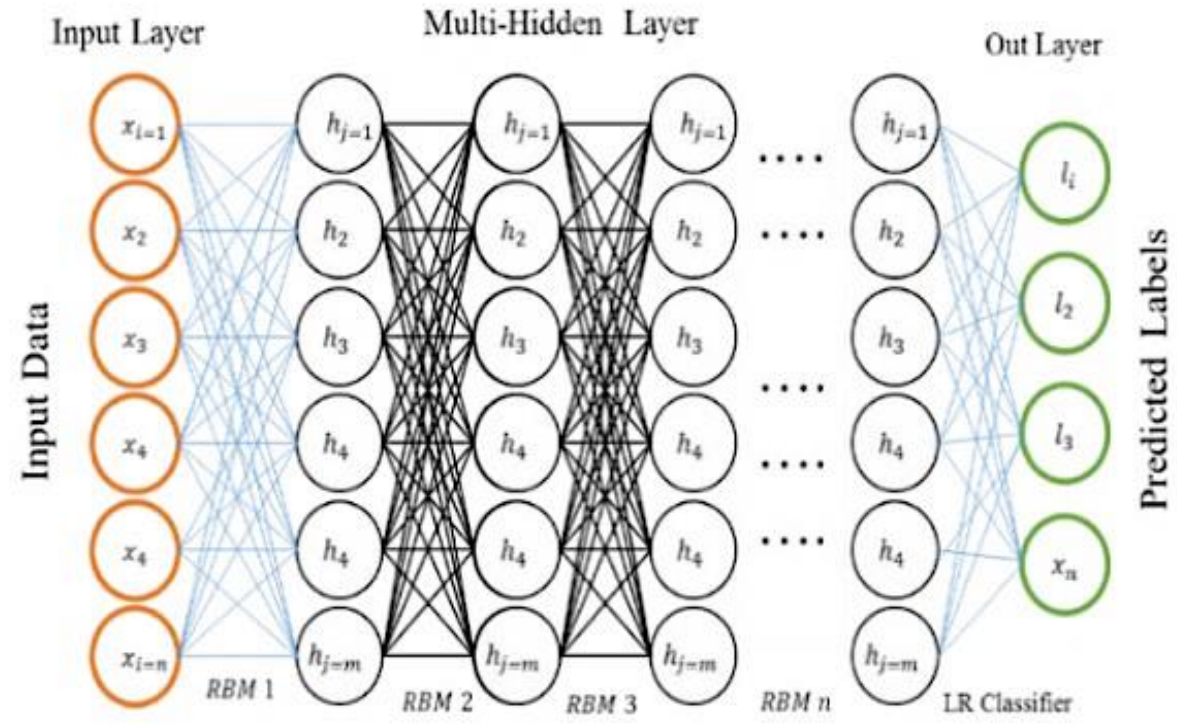

**FIGURE 10: DBN Architecture**

*C. Auto Encoder (AE)*

The AE is another unsupervised approach in DL to learn a compressed representation of raw data. This modification of an ANN has at least three layers: output, input (data), or hidden layers. An encoding function feeds encoded data to the hidden layer from the input layer. To code the compressed form of the input data, hidden layers must have fewer nodes than the input layer. The output layer plans to utilize a decoder function to decode or reconstruct the hidden layer, as depicted in Figure 11.

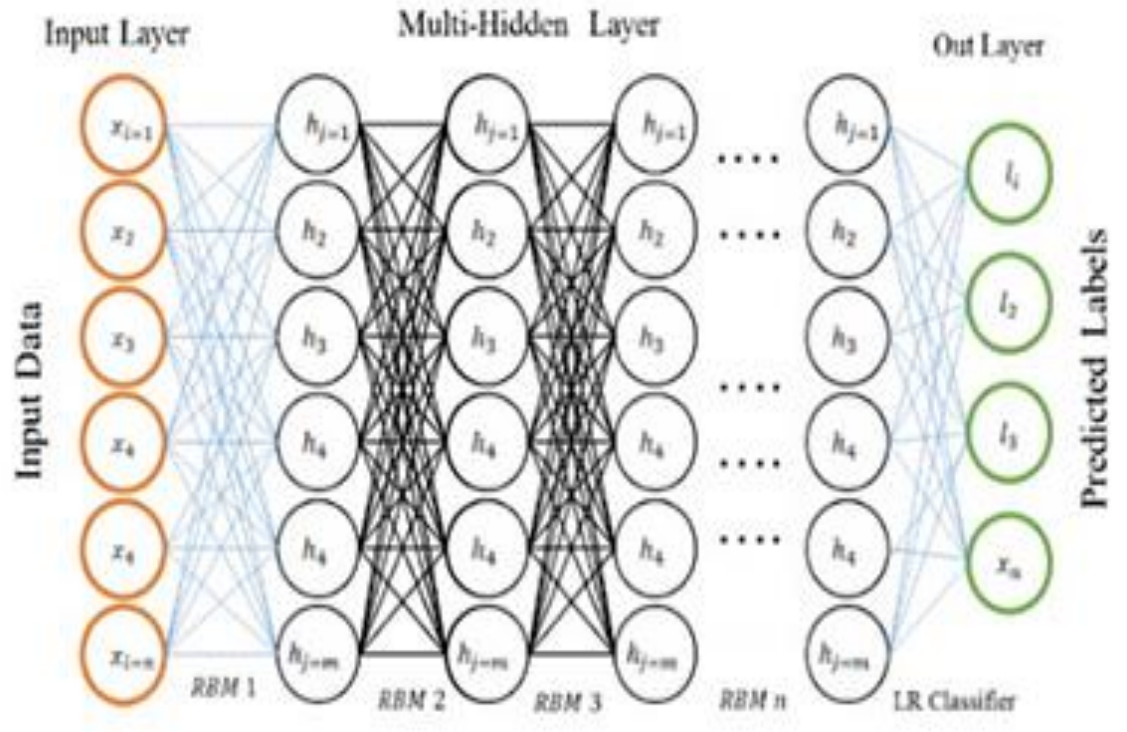

**FIGURE 11: Autoencoder Architecture**

The difference between the output and input layers is employed to create an error function, and weights are modified to minimize error. In this context, unsupervised learning learns talent parameters as weight and basis. Therefore, the traditional unsupervised learning approach is not compared with the outputs against learning the essential features. Furthermore, the versions of an AE are denoising, spare, and contractive, where the denoising type is commonly used for a NIDS and is significant with datasets in network security[58]

*D. Convolutional Neural Network (CNN)*

The CNN is a supervised deep-learning algorithm with a three-layer design: flattening, pooling, and convolutional layers. The layers are primarily used to determine the significant features to improve image processing, classification, segmentation, and other auto-correlated data [59]. A typical CNN consists of an input layer that obtains the data input and a convolution layer that generates the feature map by applying a filter matrix to the input data. The pooling layer from the convolution layer determines the feature map's significant values. The flattening layer converts learned multidimensional learned features to one dimension. This intellectual feature is fed to a fully connected layer that discovers and classifies line connections into normal or abnormal classes. Therefore, the CNN model can be implemented with the sigmoid or SoftMax function to specify a probabilistic value for each category. Remarkable CNN structures are Google Net, AlexNet, ResNe, and VGGNet [60]. A CNN's fundamental architecture for classifying outputs and processing inputs is depicted in Figure12.

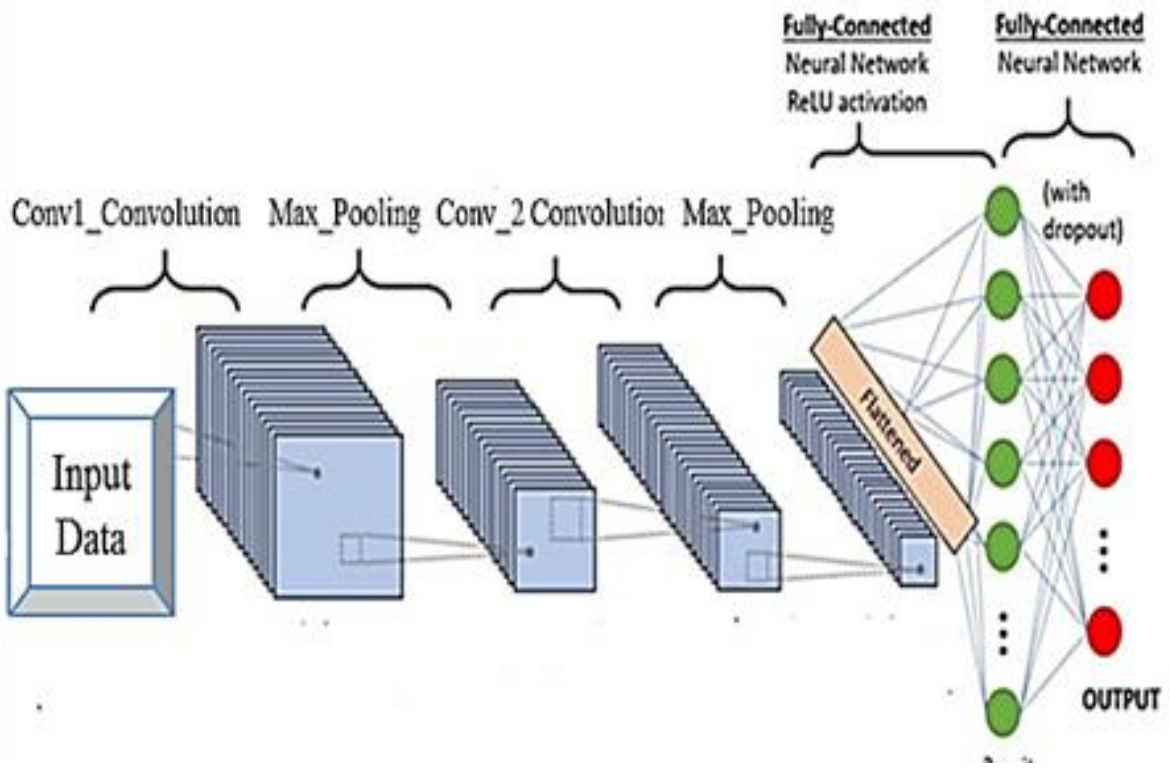

**FIGURE 12: CNN Architecture**

*E. Recurrent neural networks (RNN)*

The RNN is a supervised learning approach of a feed-forward neural network that uses the internal state known as internal memory. The RNN architecture is different from a traditional neural network. It is a correlation between units or nodes in the same layer and feedback to preceding layers [61]. RNNs may maintain a recollection of previous inputs attributable to this feedback. As an outcome, the RNN contains a lot of parameters, and training a model takes a long time.

In contrast, an RNN outperforms a traditional network by solving the vanishing gradients [62]. RNNs consist of a deep architecture. A long short-term memory (LSTM) network is what it is popularly known as. There are three gates in an LSTM network: input, forget, and output. The critical differentiator is network feedback, which might come from a combination thereof the output layer or a hidden layer [63]. In the condition of the binary and multiclass classification of the NSL-KDD dataset, RNN-IDS was recommended to investigate a variety of hidden units and learning rates. Therefore, the findings revealed that the algorithm's accuracy affects numerous learning rates and the number of hidden units. The most dependable accuracy for binary and multiclass cases was attained with 80 hidden units and a learning rate between 0.1 and 0.5.

In comparison to ML models, the proposed model performed well. The weaknesses of this research are that it requires high computational processing and takes a long training time with a low detection for skewed attacks such as R2L and U2R attacks. For the NIDS model with multi-layer perceptron (MLP) and SoftMax module, the gated recurrent unit (GRU) technique was devised. The proposed IDS model consists of preprocessing and a classifier. The essential features are extracted and stored by the GRU. The information is transmitted to the MPL output layer, classifying the line connection as abnormal or usual. The result appeared to outperform LSTM in terms of detection rate and fast training as fast calculation, and it is easy to stack [64].

## V. DATASET

For the algorithm to learn, a data set (or dataset) is a collection of data utilized to generate by entering a set of training data for which the classes are pre-labeled. Here are the details of the public data utilized by the researchers for experimenting with performing their intended works, namely KDD CUP 99, NSL-KDD, UNSW-NB15, CICIDS2017, CICIDS2018, and ToN_IoT (refer to Figure 13). Therefore, a report of the dataset and the attacks were discussed in the following sub-section.

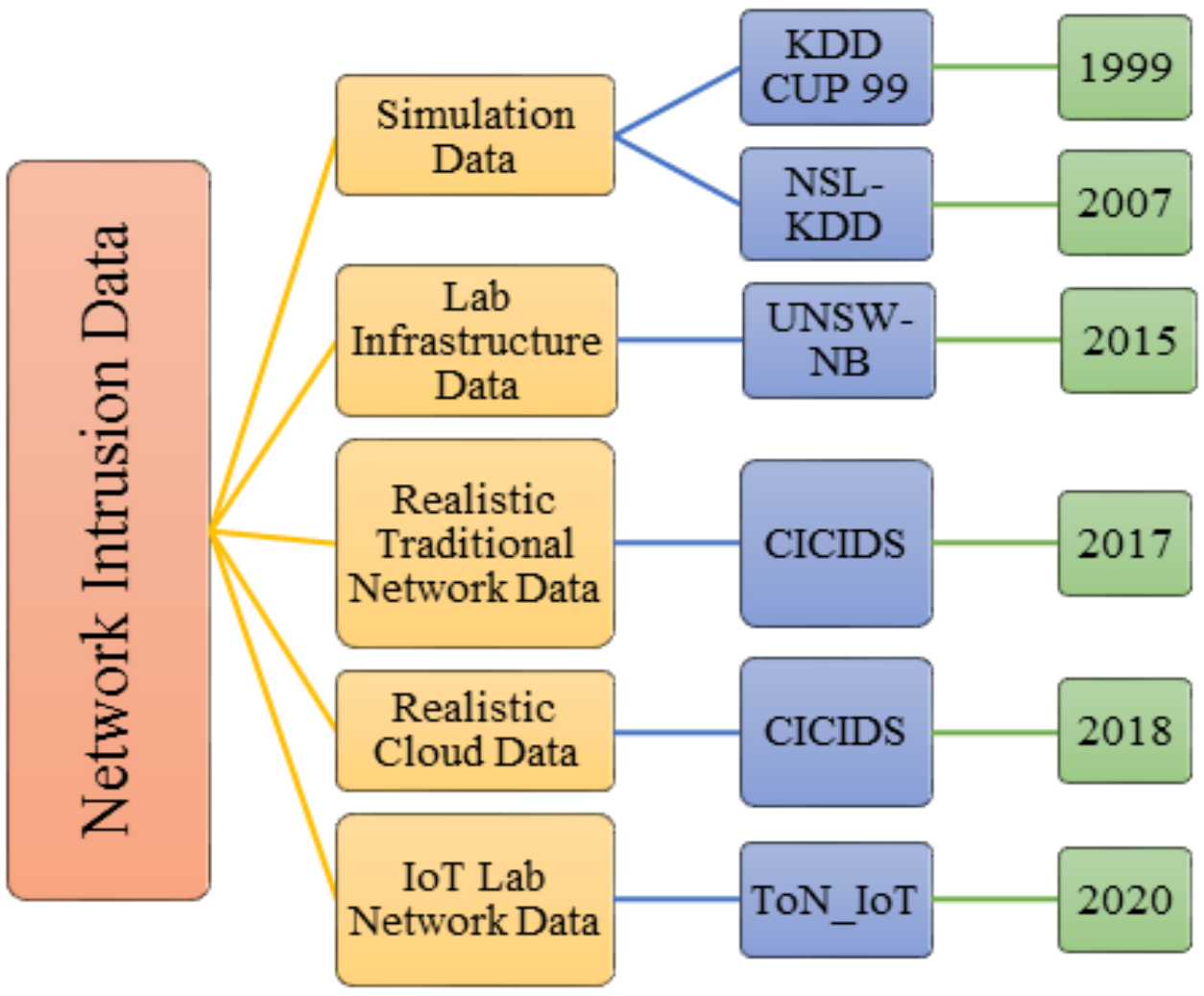

**FIGURE 13: Data Network**

### A. KDD CUP 99

This dataset is designed to assess an IDS constructed in 1998 by MIT Lincoln Labs as a simulation dataset. KDD CUP 99 is intensively applied in the AI field as it consists of two parts (training and test data) and consists of 41 features representing characteristics and content traffic [65]. The standard dataset of KDD CUP 99 consists of approximately five million raw data. An estimated eighty percent of this dataset has attack data. The dataset is categorized into two main categories: attack and routine.

**Normal:** non-attack or expected behavior of data.

**Attack types:** denial of service (DOS), user to root (U2R), probing attacks (Probe), and root to local (R2L).

There are twenty-two attack types, and every single one is part of an attack category classified above. The KDD CUP 99 datasets contain features shown as an actual number and text values about request categories. Furthermore, these datasets also consist of an addend characteristic at the end of the process, indicating the data labelling segregating an intrusion or normal category. The KDD Cup 99 dataset comprises 41 extracted features that capture different aspects of network connections, including basic information, content-related characteristics, and traffic properties. These features serve as inputs for developing and evaluating machine learning models for network intrusion detection. Basic features provide basic information about network connections. They include attributes such as duration (the length of the connection in seconds), protocol type (TCP, UDP), service (HTTP, FTP), and source/destination IP addresses and port numbers. Content-Related features capture characteristics of the network packets or data content within the connections. They include attributes such as the number of failed login attempts, the number of shell prompts, the number of urgent packets, and the number of bytes transferred. Traffic features describe the statistical properties of network traffic. They include attributes such as the number of connections to the same host in the past two seconds, the number of connections to the same service in the past two seconds, and various other statistical measures related to time-based properties of network traffic.

### B. NSL-KDD

This dataset has been proposed as a solution to some of the difficulties with the KDD'99 datasets. It includes an updated version of the whole KDD CUP 99 dataset. It possesses data features with a different number of classes [10]. Developing the NSL-KDD dataset by reducing the data size by deleting replicated records facilitates machine learning algorithms' performance. In the NSL-KDD dataset, there are a total of 41 extracted features, similar to the original KDD Cup 99 dataset. However, the NSL-KDD dataset introduces some modifications and improvements to enhance its usability and address certain issues in the original dataset. The NSL-KDD dataset comprises 41 extracted features, including basic features, content-related features, traffic features, and binary features. Binary features were introduced in the NSL-KDD dataset by introducing a new group of binary features that encode specific aspects of network connections. These binary features represent the absence or presence of certain attributes, such as whether a particular type of attack was present or not in a connection. This addition allows for a more precise and balanced representation of different attack categories.

### C. UNSW-NB15

This dataset is a network intrusion dataset. The IXIA Perfect Storm tool software generated this dataset. Unlike NSL-KDD, it includes primary variants of different ID cases appearing more frequently nowadays. The number of regular classes is 175,341, and there are 82,332 anomaly classes [42]. The attacks are worms, shellcode, reconnaissance, generic, exploits, DoS, backdoors, analysis, and worms. The UNSW-NB15 dataset consists of 49 extracted features capturing various aspects of network traffic, including basic information, content-related characteristics, statistical properties, connection behavior, time-related factors, and service-related attributes. These features are utilized to build and assess machine learning models for network intrusion detection. Basic features provide fundamental information about network connections. They include attributes such as source/destination IP addresses, source/destination port numbers, protocol type (TCP, UDP), and flags (SYN, ACK) associated with the connection. Content-Related features capture characteristics of the network packets or data content within the connections. They include attributes such as the number of bytes transmitted in both directions, the duration of the connection, and the number of packets exchanged. Statistical features describe the statistical properties of network traffic flows. They include attributes such as the average time between packets, the standard deviation of packet length, and the rate of incoming and outgoing packets. Connection-Based features focus on the behavior of network connections. They include attributes such as the number of connections from the same source IP address, the number of connections to the same destination port, and the ratio of incoming to outgoing packets. Time-Based features capture time-related characteristics of network traffic. They include attributes such as the time since the start of the first connection in seconds, the duration of the connection relative to the total duration of the data capture, and the time difference between connections. Service-Based features pertain to the specific network services used in the connections. They include attributes such as the service name (HTTP, FTP), the number of connections associated with the service, and the ratio of connections for each service.

### D. CICIDS2017

CICIDS2017 has approximately simulates real-world networks. The CICFlowmeter-V3.0 is intended to create realistic data that includes a set of labels and features. The dataset includes email protocols, SSH, FTP, HTTPS, and HTTP transmitted via an entire network of twenty-five nodes [66]. The data is collected over a duration of time. The following attacks are recommended in the 2016 McAfee report: DDoS, Botnet, Infiltration, Web Attack, Heartbleed, and DoS. The early dataset [41] did not reveal any harmful attacks. Utilizing the B-Profile system, the dataset is employed to accomplish conceptual feature profiling of users' communication, while the Alpha profile is intended to carry out various attack scenarios. The dataset has characteristics that distinguish it from other data in terms of realism. The phenomenon of realism is covered by 11 criteria, such as completed traffic, protocols., to ensure the quality of the evaluation [4]. The CICIDS2017 dataset consists of 78 extracted features capturing various aspects of network traffic, including basic information, content-related characteristics, statistical properties, connection behavior, time-related factors, host-based attributes, and traffic-based insights. These features serve as inputs for building and evaluating machine learning models for network intrusion detection and cybersecurity research. Basic features provide fundamental information about network connections. They include attributes such as source/destination IP addresses, source/destination port numbers, protocol type (TCP, UDP), and flags associated with the connection. Content-Related features capture characteristics of the network packets or data content within the connections. They include attributes such as the number of bytes transmitted in both directions, the duration of the connection, and the number of packets exchanged. Statistical features describe the statistical properties of network traffic flows. They include attributes such as the average and standard deviation of packet length, the average and standard deviation of packet inter-arrival time, and various statistical measures related to payload and flow properties. Connection-Based features focus on the behavior of network connections[67] . They include attributes such as the number of connections from the same source IP address, the number of connections to the same destination port, and various ratios and counts related to connection behavior. Time-Based features capture time-related characteristics of network traffic. They include attributes such as the time since the start of the first connection in seconds, the duration of the connection relative to the total duration of the data capture, and various time-related statistics. Host-Based features represent information related to the host or device involved in the network connection. They include attributes such as the number of connections from the same host, the number of different services used by the host, and various ratios and counts related to host behavior. Traffic-Based features provide insights into the overall network traffic characteristics. They include attributes such as the total number of packets, the total number of bytes, and various traffic-related statistics.

### E. CICIDS2018

It was a collaborative project between the Communications Security Establishment and the Canadian Institute for Cybersecurity [68]. It provided 10 days of traffic, from Wednesday, February 14, 2018, to Friday, March 2, 2018, focusing on Amazon Web Services (AWS). This included seven different attack scenarios deemed similar to CICIDS2017: brute-force (FTP-Patator and SSH-Patator), Denial of Service (slowloris, SlowHTTPTest, Hulk, GoldenEye), Heartbleed, web attacks (Damn Vulnerable Web App, XSS,brute-force), infiltration of the network from inside, botnet, and Distributed Denial of Service with port scanning [69]. However, CSE-CIC-IDS2018 was a more complete dataset than CICIDS2017, with more data and different network topologies. The CICIDS2018 dataset consists of 79 extracted features capturing various aspects of network traffic, including basic information, content-related characteristics, statistical properties, connection behavior, time-related factors, host-based attributes, and traffic-based insights.

### F. IoT Dataset

The TON_IoT datasets are new generations of Internet of Things (IoT) and Industrial IoT (IIoT) datasets for evaluating the fidelity and efficiency of different cybersecurity applications based on Artificial Intelligence (AI). The datasets have been called ‘ToN_IoT’ as they include heterogeneous data sources collected from Telemetry datasets of IoT and IIoT sensors, Operating systems datasets of Windows 7 and 10, and Ubuntu 14 and 18 TLS and Network traffic datasets. The datasets were collected from a realistic and large-scale network designed at the IoT Lab of the UNSW Canberra Cyber, the School of Engineering and Information technology (SEIT), and UNSW Canberra at the Australian Defence Force Academy (ADFA). The datasets were gathered in parallel processing to collect several regular and cyber-attack events from IoT networks. A new testbed was developed at the IoT lab to connect many virtual machines, physical systems, hacking platforms, cloud and fog platforms, IoT and IIoT sensors to mimic the complexity and scalability of industrial IoT and Industry 4.0 networks [70]. For example, IoT/IIoT datasets typically capture various aspects of sensor data, network communication, and device behavior in connected environments. The extracted features can be categorized into several types, including sensor data features, network communication features, device metadata features, Time-Based features, contextual features, and derived features. Sensor data features represent measurements or readings from IoT/IIoT sensors. They may include attributes such as temperature, humidity, pressure, light intensity, vibration, sound level, or any other relevant environmental or physical measurements. Network communication features capture information related to network communication between IoT/IIoT devices. They may include attributes such as source/destination IP addresses, source/destination port numbers, protocol type (e.g., MQTT, CoAP), message payload size, message frequency, or other network-related characteristics. Device metadata features provide descriptive information about the IoT/IIoT devices themselves. They may include attributes such as device ID, device type, firmware version, hardware specifications, location, or any other relevant device metadata. Time-Based features capture time-related characteristics of IoT/IIoT data. They may include attributes such as timestamps, time intervals between data points, or any temporal patterns or trends within the dataset. Contextual features capture contextual information associated with the IoT/IIoT data. They may include attributes such as location information, user context, environmental conditions, or any other contextual factors that can provide additional insights into the dataset. Derived features are calculated or derived from the raw sensor or network data. They may include statistical measures such as mean, standard deviation, maximum, minimum, or more complex derived features such as frequency domain analysis, waveform characteristics, or pattern recognition. The number of extracted features in IoT/IIoT datasets can vary significantly depending on the specific dataset and the nature of the collected data. It can range from a few dozen to hundreds or more features, depending on the complexity of the IoT/IIoT system and the goals of the dataset.

## VI. VALIDATION METHODOLOGY, RECENT TRENDS AND OBSERVATIONS

To be efficient and validate the performance of anomaly NIDS models, they must be professionally built using a significate AI algorithm and up-to-date dataset for detecting zero-day attacks. Five factors can affect the validity of deep learning methodology, factors are definition rules (network dataset), efficient tools (algorithms), and time detection as a crucial point due to being resource-intensive in contexts of time consumption and computational resources. Furthermore, a type of classification task is another important factor that affects building the anomaly NIDS. Different types of classification tasks in deep and machine learning, there are two main classification tasks in machine learning: binary or multi-class for imbalanced classifications. In general, it is difficult to build multi-detection attacks due to being high false positives, whereas highly accurate and easy to detect one attack. According to those hypotheses, the benchmarking assessment of the anomaly NIDSs that are depended on the used algorithms, their effectiveness criteria, datasets, classification task, and time detection. We are using a meta-analysis technique to collect the data comparison or values numbers of evaluation factors from recent research as shown in Appendix A, we have observed from those factors as follows: -

## A. ANOMALY-NIDS ALGORITHMS

According to the studied literature, writers have proposed the IDS system utilizing the DL technique in the recent three years, as demonstrated in Figure 14. It can be noticed that 50% of DL approaches were available to construct the IDS model. On the other hand, only 20% of suggested models employ a hybrid approach that combines DL and ML techniques, whereas only 30% of recommended solutions depend on ML approaches. As previously said, DL algorithms are complex and demand a lot of computing power. In addition, Figure 15 reveals how frequently writers use DL algorithms to create successful IDS models. The four standard multiple algorithms employed for NIDS are RNN, CNN, DNN, and AE, all of which are DL techniques. Subsequently, ML techniques, for instance, SVM and RF, were presented as a hybrid strategy to enhance DL algorithms. Different metrics are typically used to measure anomaly detection in NIDS models. The effectiveness of ML/DL reflects performance based on accuracy, confusion matrix, precision, recall, and f-score.

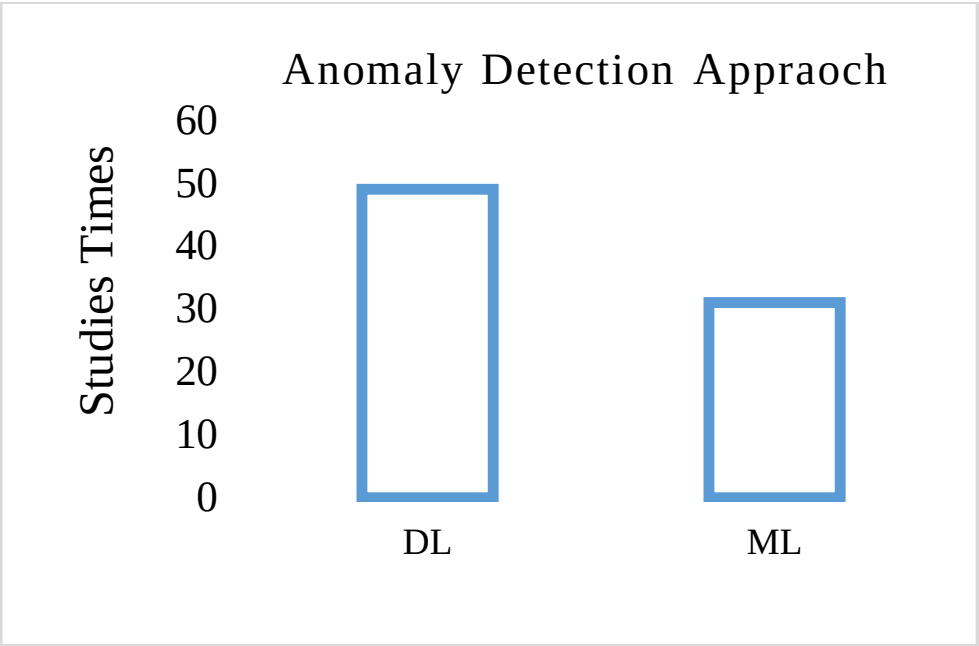

**FIGURE 14: Frequency of ML and DL Algorithms Used in the Surveyed Materials**

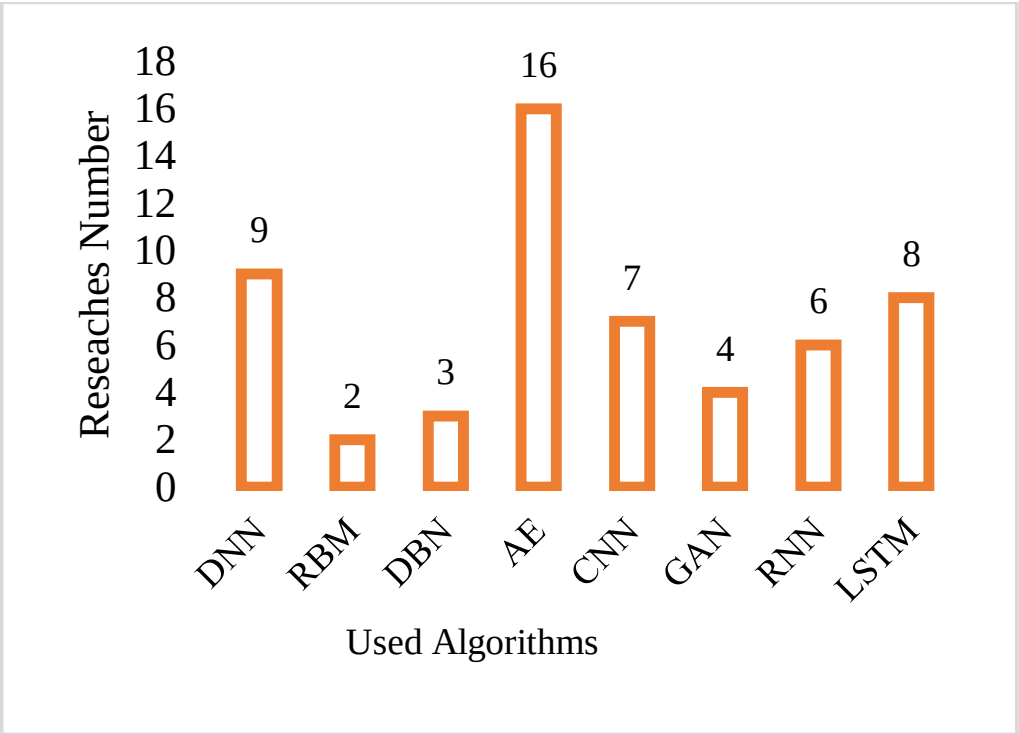

**FIGURE15: Frequency of DL Techniques Used in the Surveyed Materials.**

Figure 15 show the reviewed articles in the group distribution for DL approaches based on summarization deep learning works in appendix A. It can be observed a AE algorithm has used more than supervised or sim-supervised learning algorithms to build NIDS models, The AE algorithm had most used in a NIDS model with 16 times such as [64,65, 87, 91]. While DNN algorithm had used to detect attacks by some authors such as [52,90,95,97] with 9 time. With unsupervised and hybrid learning types, DBN and AE algorithms and their different versions had the most widely used algorithms for proposing NIDS solutions. The majority of the time, AE is expressly embraced and utilized for feature reduction and extraction. Meanwhile, for classification purposes, it is accompanied by the ML-based classifier. The SVM, RF, DT, and K-means build NIDS models is used more often than other conventional ML, as shown in Figure 16.

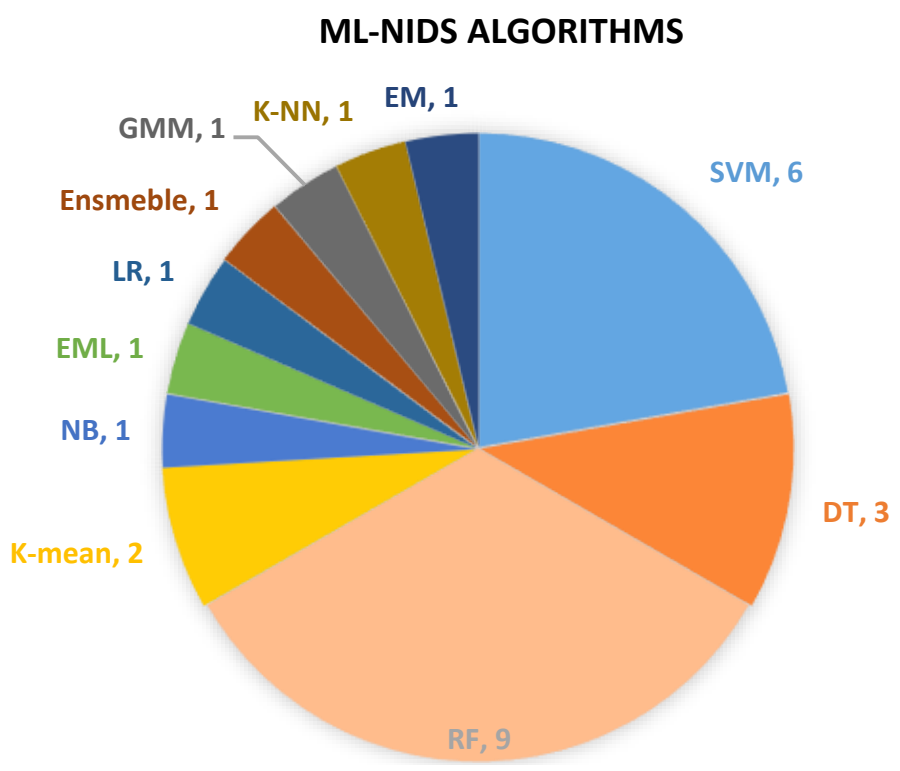

**FIGURE 16: Conventional ML Algorithms for NIDS**

## B. Network Intrusion Dataset

Most datasets are designed to provide to researchers by different institutes of cybersecurity based on the infrastructure network such as cloud, Software Defined Network, or Web database server, the researchers are utilizing data traffic to build an anomaly NIDS, as shown in Figure 17 and Figure 18. AL-IDS models had built and validated using MIT simulation data. MIT simulation data includes two public data, KDD CUP 99 and NSL-KDD utilizing (54%). Another data network is generated by the local infrastructure network using UNSW_Lab with (15%) of usage data by recent researchers for building the anomaly NIDS models. Modern network design is completely unconventional forms that are not the same as an infrastructure network in the last age, and these datasets are ancient. It becomes more dangerous to build the production tool for the advanced networks, and there might be more possibilities that the suggested methods will not play well when placed in real-world situations. Advanced datasets, for instance, BoT-IoT 2018 and CICIDS 2017, are utilized for training and evaluating the AI-IDS models, which will appear more significant in the actual world than models that utilize an outdated dataset. Unfortunately, few types of research, as 31% of studies propose a realistic dataset for training anomaly-NIDS, as shown in Figure 17.

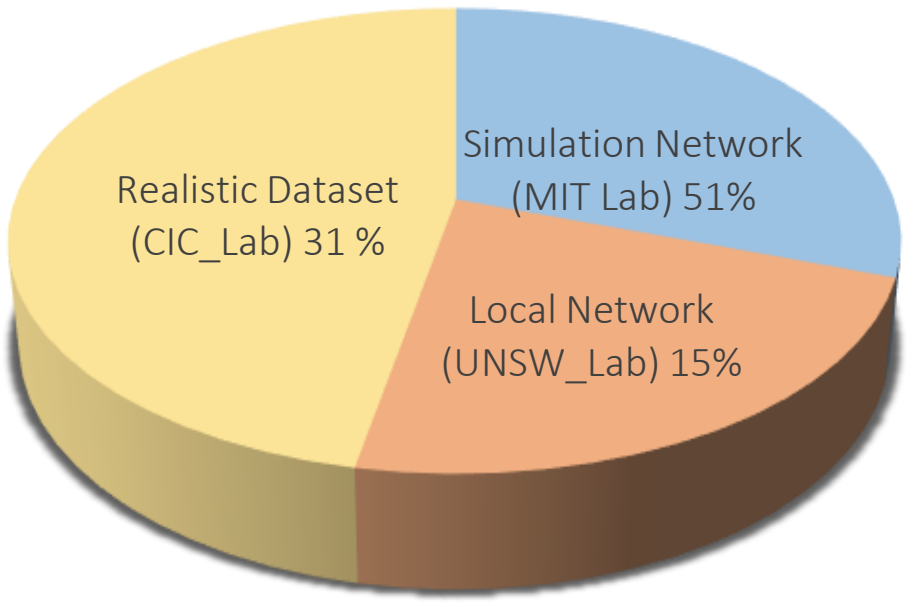

**FIGURE 17: Percentage of Network Intrusion Dataset**

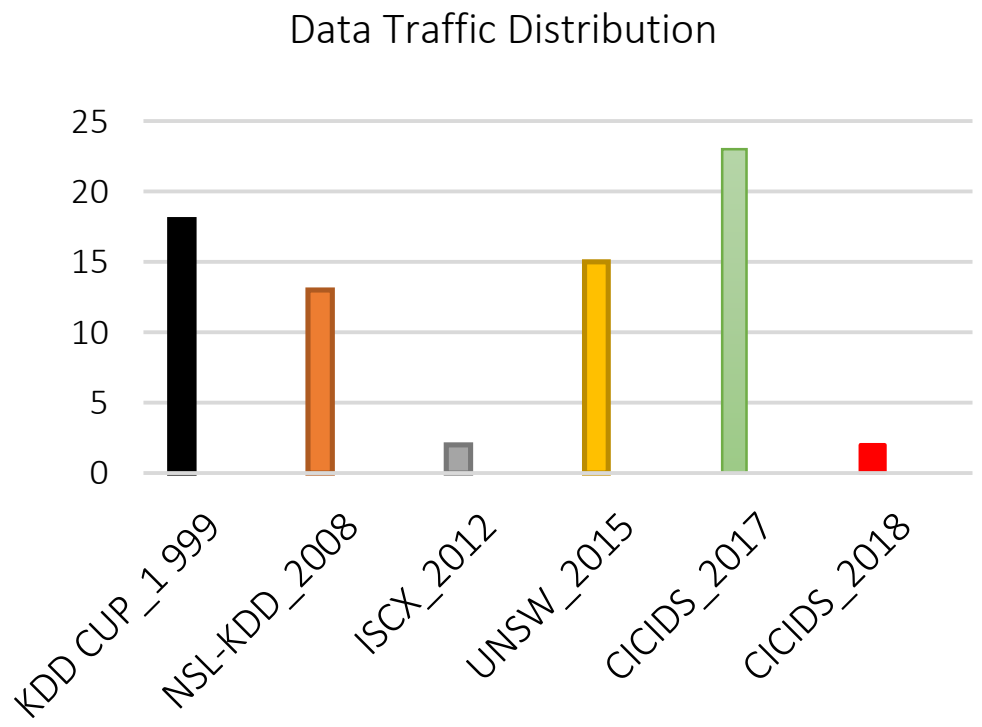

**FIGURE 18: Frequency of Proposed Dataset**

## C. Classification Task

The classification is an essential task concerning the ML or DL approach. Two parts to training the anomaly models to be an IDS classifier are binary and multi-task. A binary task is a simple operation to classify connection lines into normal and abnormal classes. At the same time, multi-task is a complex operation to classify data into multi classes as usual or attacks. Table 3 shows a statistical classification task for anomaly NIDS detection, the classification task has been proposed to build anomaly NIDS around (47 rates) in both binary and multi-classification.

**TABLE 3: STATISTICAL CLASSIFICATION TASK IN APPENDIX A**

| Classification Task Type | Statistical Studies |
|---|---|
| Muli- Classification | 47 |
| Binary-Classification | 47 |

The AI algorithms can efficiently train and fit parameters for building anomaly-IDS models in binary classification tasks. At the same time, the multi-classification ML/DL models have many parameters that need to improve or optimized for concluding optimal nonlinear equations for anomaly-IDS models.

### *Binary Classification*

Figures 19 and 20 show the data comparison to the overall and per-class measurement for anomaly NIDS models in binary classification. Overall measurement shows the anomaly detection system's ability to detect attacks in binary form while ignoring imbalance or realistic features. The accuracy and precision are higher used by recent researchers than other metrics (Recall and F-score).

The specific measurement is used to show the performance of NIDS to detect per attack or class regarding the imbalance dataset. The precision, recall, f-score, and accuracy were used with different percentages as research metrics. The recall and f-score were used higher to check the models' fitting than other metrics (precision and f-score).

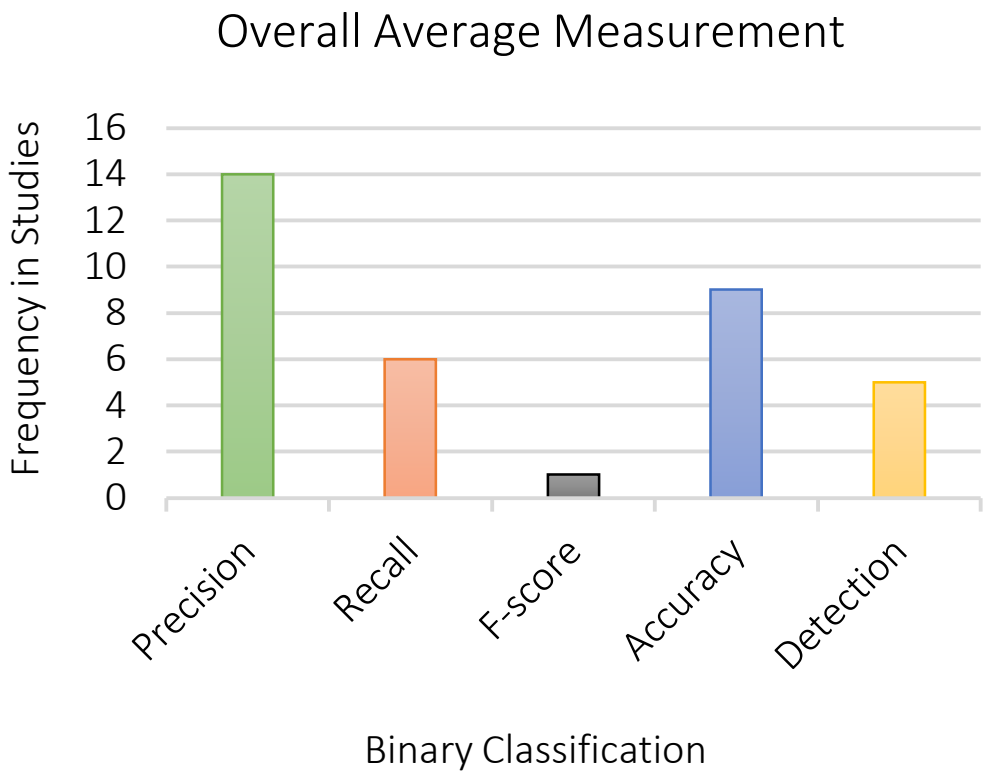

**FIGURE 19: Overall Measurement in Binary Task**

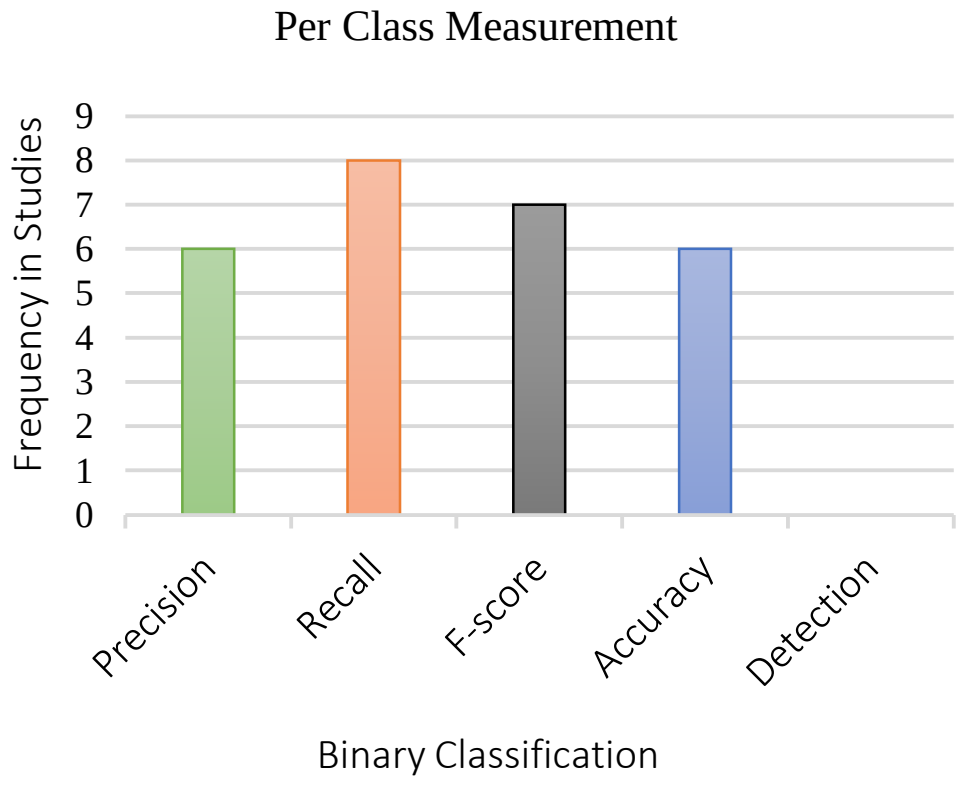

**FIGURE 20: Per class Measurement in Binary Task**

### *Multi-Classification*

Most cybersecurity datasets are multi-classes and imbalance labels. Building an anomaly detection system based on binary tasks and costly detection with only one attack and a regular connection is unreasonable. Multi-classification is a problematic task via anomaly network detection, the researchers are required to find an optimal non-linear question for building the AI models for detecting many attacks simultaneously.

Overall and per class is used to evaluate the system, the researchers used average or overall measurement for all samples or line connection using precision, recall, f-score, accuracy, and detection to show researchers' contributions to the anomaly intrusion detection system. As seen in Figure 21, the overall frequency is more accurate than the rest measurement, followed by the precision rate. To be a more specific evaluation of anomaly NIDS, the authors proposed metrics to show the detection of per-class or attack, as shown in Figure 22. The per-class measurements are different levels that have been used to evaluate the detection of each attack in many types of research, the accuracy is used more than other metrics to measure the accuracy the anomaly detection per class.

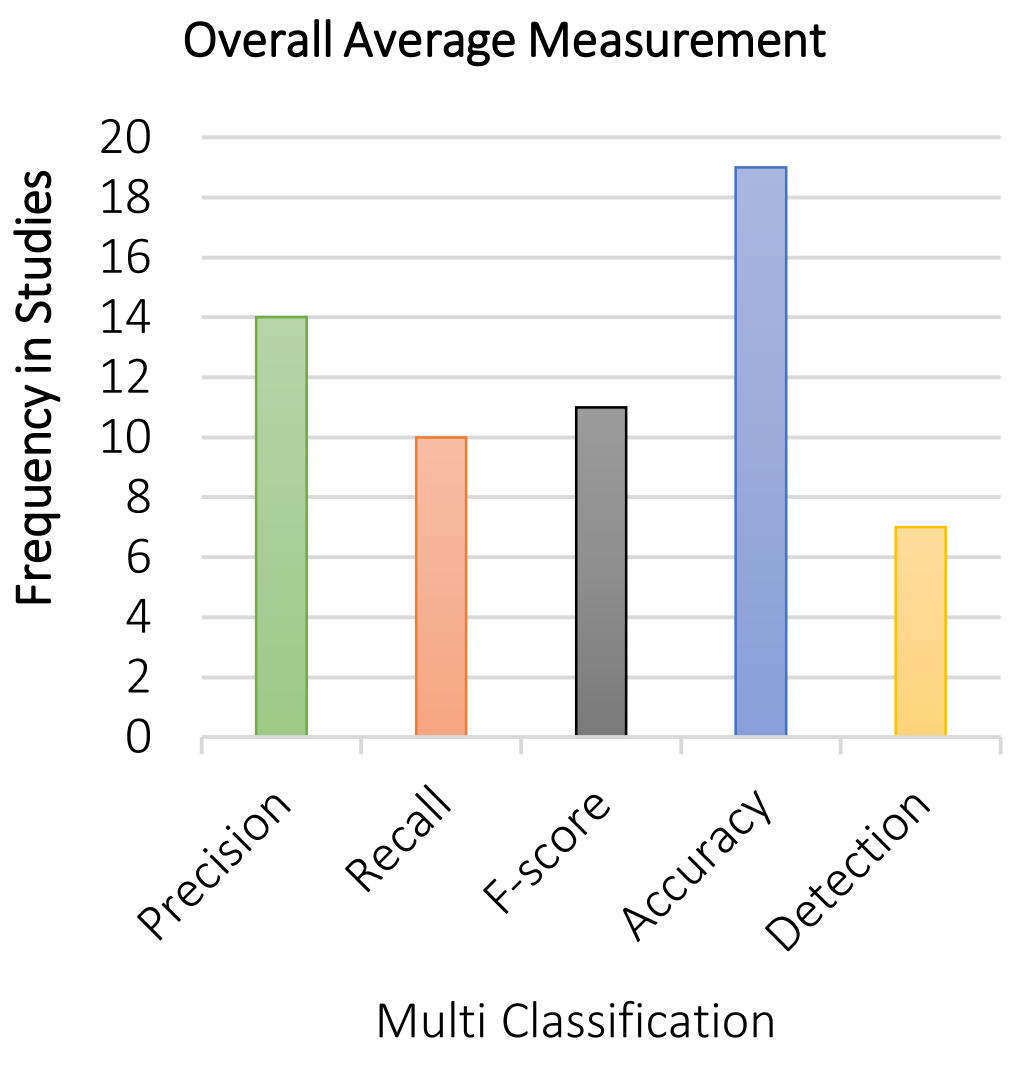

**FIGURE 21: Overall Measurement in Multi-Class Task**

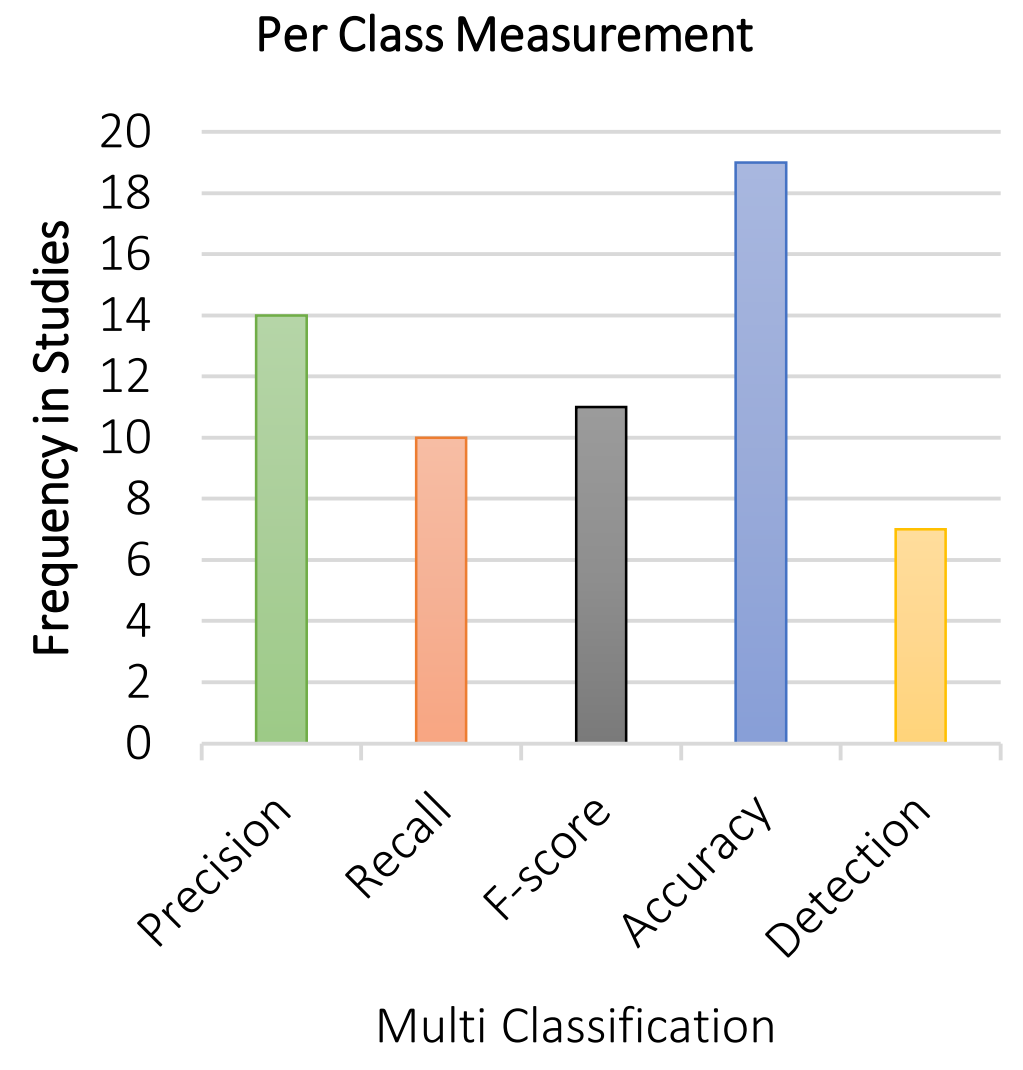

**FIGURE 22: Per class measurement in multi-class task**

### D. Anomaly-NIDS Effectiveness

The NIDS effectiveness is measured through confusion metrics of cybersecurity, which are accuracy, precision, and recall. Other authors could propose time detection or prediction to measure the ability of efficiency as shown in Appendix A. This article is contained a lot of papers to anomaly detection. The best conditions for the anomaly NIDS effectiveness, that is pointed to optimal detection of attacks and the benchmarked works indicate actual effectiveness that had been selected from appendix a. We discuss those conditions to give a clear image of the actual performance of anomaly NIDS based on high values for each accuracy, precision, recall, and f-score.

Most authors had proposed couples or a few metrics to measure the effectiveness of AI algorithms such as accuracy. The accuracy had only used to measure the effectiveness of DL-NIDS models, and the results of the accuracy rate could range between 0.83 to 1.00 with (KDD CUP 99, UNSW-NB15, CICIDS2017, and CSE-CIC-IDS2018) datasets, respectively [47], [54], [60-66], [76], [121], [122], [124-126]. Similarly, the precision rate had used to measure the effectiveness of anomaly detection but examination of the connection lines to be true and false actives by different AI algorithms. Where the FR algorithm had performed multi-detection or classification attacks based on KDD CUP 99 data traffic, the precision rate had been indicated to acceptable level at 0.99, 0.92, 0.17, 0.66, 0.55 for Normal, and four types of attacks such as DoS, U2R, and R2L, respectively [49]. The authors used a simple way to find overall precision rate by calculating the average true positive and negative for five classes with a 0.99 rate by the AE algorithm [82]. The CNN algorithm had been proposed to improve the detection of multi-classes (regular and four types of attack) based on the NSL-KDD dataset with a good precision percentage of 1.00 (regular), 1.00 (DoS), 0.70 (U2R), 0.70 (R2L) and 1.00 (Probe) [83]. The precision used to measure the overall average effect of an anomaly NIDS by different algorithms such as AE, RF, RBM, and DNN, the authors used to train and evaluate the anomaly NIDS based different datasets is NSL-KDD, AWID, UNSW-NB15, CIDDS-001, and Realistic data [85], [87], [89], [90], [92], [105], [108-109]. The excellent precision had performed by the RF algorithm with around a 1.00 rate for normal lines and other types of attacks of the NSL-KDD dataset [110].

Whenever solutions for solving a false positive and false negative together in one problem in cybersecurity, it must use full metrics: precision, recall, f-score, and accuracy to ensure the result or evaluation is true. It is clear from Appendix A that contains a few studies [52], [88], [122], and [74], which observed to increase the effectiveness of anomaly NIDS by solving false positive and false negative issues in the imbalanced dataset (rare attacks and many samples of a regular action). Indeed, the authors' results are nearest to realistic as well as benchmarking works due to including full or some of metrics to the ability of anomaly NIDS model by precision to false negative, recall to false positive, and f-score to balance between precision and recall.

In KDD CUP 99, normal actives and attacks (DoS and Probe) which had detected by a k-mean algorithm with acceptable levels as well as low detection for minor samples of U2R and R2L attack [52]. The effectiveness had enhanced by using the AE algorithm to detect connection lines whether normal or attacks (DoS, U2R, and Probe) in different types of data traffic (NSL-KDD), but anomaly detection could be a low effective for detecting R2L attacks [88] as shown in Figure 23. The effectiveness of anomaly detection had measured by three metrics precision, recall, and f-score using LAN data traffic (UNSW-NB15) as shown in Figure 24. Low effectiveness for detecting DoS, Analysis, Backdoor, and Shellcode [71].

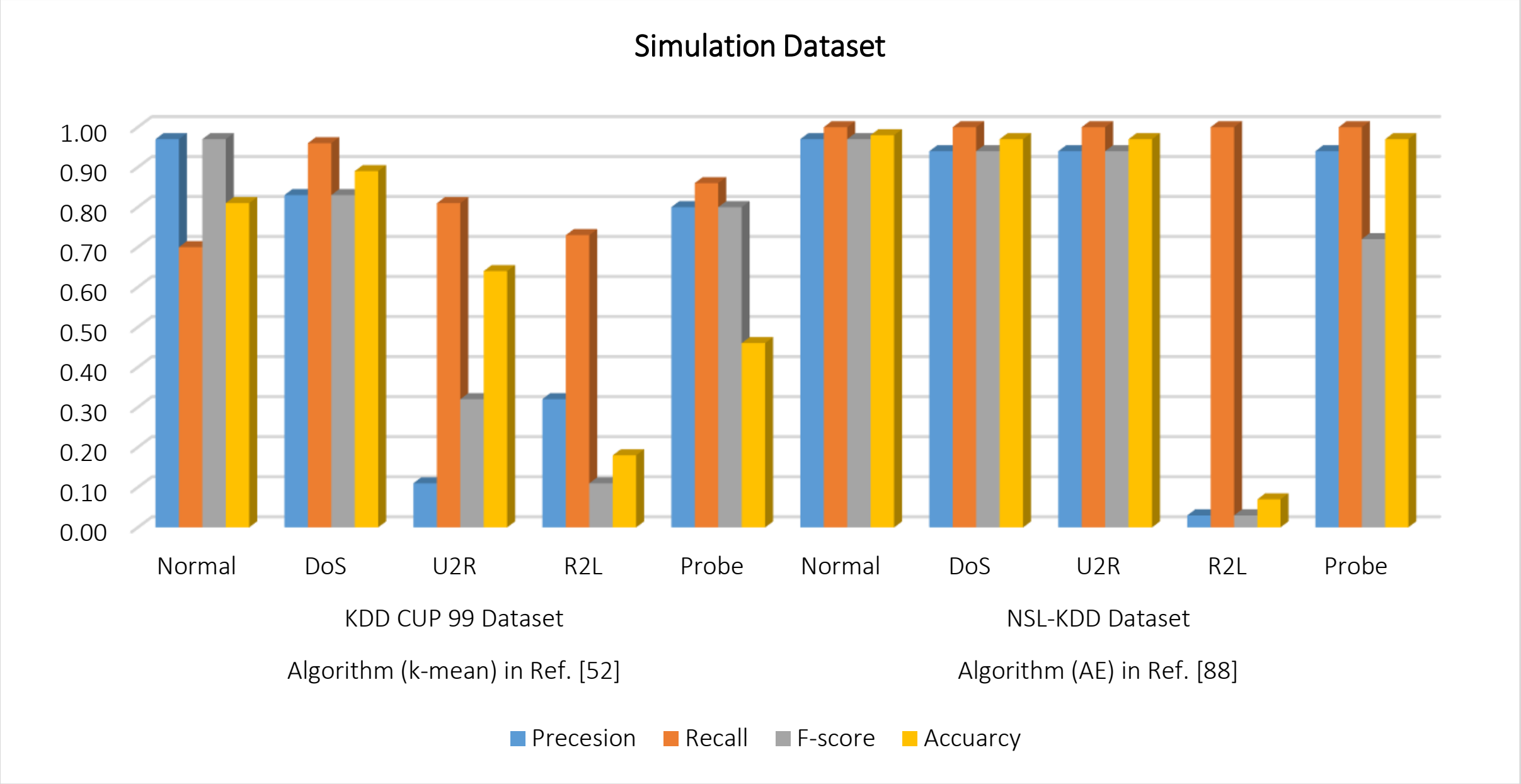

**FIGURE 23: Effectiveness of anomaly NIDS based (KDD CUP 99 and NSL-KDD)**

These metrics are roughly by starting from 0.10 to 0.40 rate. Opposite case with other connection lines, the effectiveness is increased detection for Normal, Generic, Exploits, Fuzzers, Recon and Worms with precision, recall and f-score by starting from 0.70 to 1.00 percentages [74].

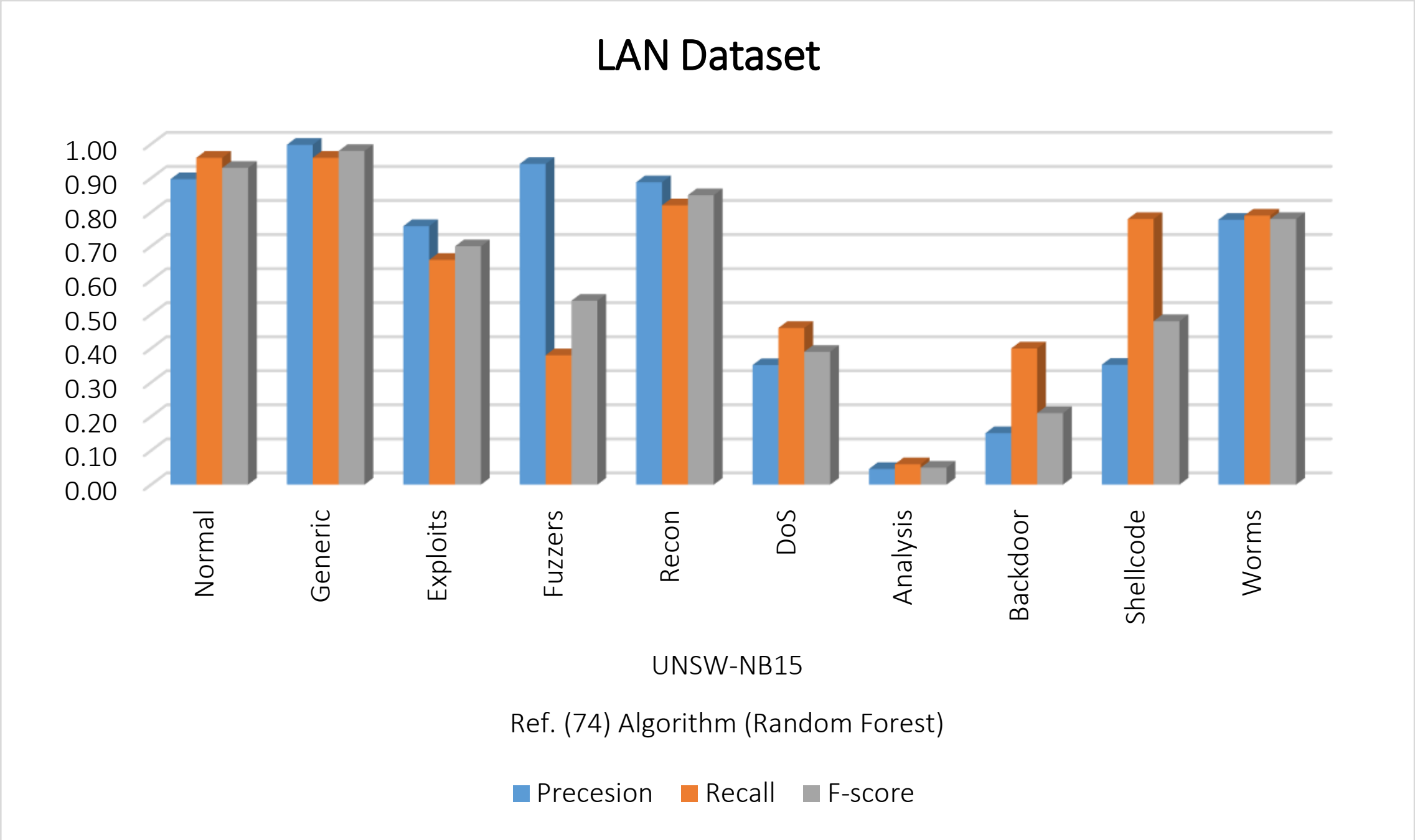

**FIGURE 24: Effectiveness of anomaly NIDS based (UNSW-NB15)**

In addition, the effectiveness of the AE algorithm had evaluated using a realistic network (CICIDS2017) which contains modern attacks as shown in Figure 25. The result had shown three metrics (precision, recall, and f-score), the effectiveness of performance is high detection for whole connection lines except of low detection in filtration and botnet attacks [122].

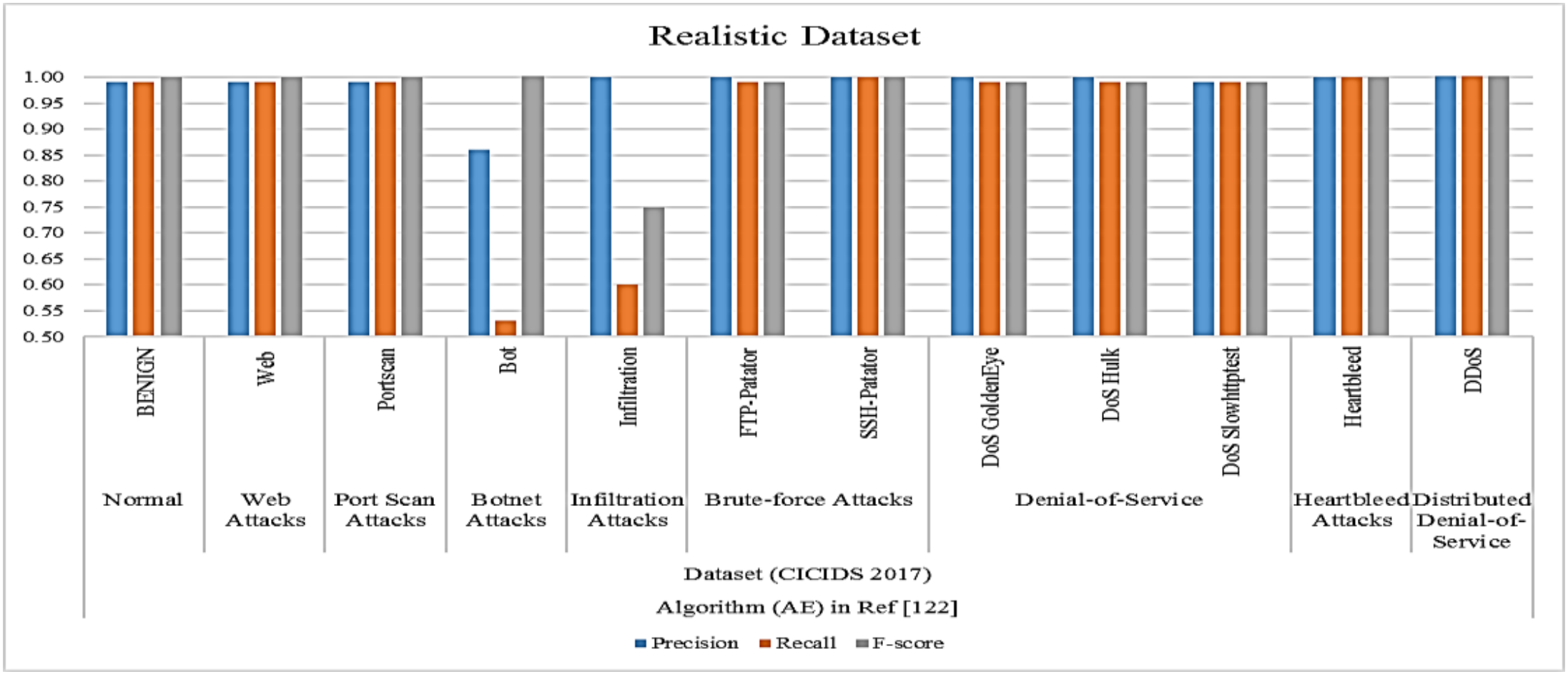

**FIGURE 25: Effectiveness of anomaly NIDS based (CICIDS2017)**

## VII. Current Challenges of Research

The current challenges are summarized and highlighted for viewing directly for the scenarios of NIDS articles to the researchers about the high-security issues occurring today as follows:

### A. False-Positive Issue

Most suggested IDS approaches exhibit false negatives for low-frequency attacks, according to the existing findings. Studies find this drawback with multi-task classification methods, which will be inefficient for detecting attacks with fewer samples during the training dataset [65, 66] .These issues are caused by an imbalance class problem, which biases the model's training for the majority samples of the class compared to minority attack classes.

### B. Low-quality dataset

The utilization of a suitable dataset is an important goal in the development of a DL-based IDS. The existing suggested DL-based IDSs do not deliver dependable overall performance, as detailed in Section 4. They utilize the KDD CUP 99 or NSL-KDD benchmark datasets, which feature ancient traffic, do not reflect existing attack scenarios or traffic habits, and have real-time features. Assessing more existing datasets, for instance, the CICIDS2017 IDS dataset [73], [76], [83], [101], and the Bot-IoT dataset, can address this issue by acquiring traffic from simulated settings. Datasets can also be created, and published datasets for many areas, for instance, industrial control systems (ICS) and IoT, are accessible as well as SDN [74].

### C. No Accurate-Time Detection

An IDS's performance in a real-world environment is another research problem. Neither approach is built in a real-time setting since most NIDS models are trained and evaluated utilizing a public dataset. Therefore, the NIDS will generate a lot of false alarms, and it is yet unclear how real-world scenarios with noise values would function. As previously indicated, the majority of them continue to depend on obsolete datasets for testing. Furthermore, the real-time performance of NIDS is not evaluated utilizing a destructive test. Thus, the suggested methodology's most significant difficulty is to be effective in real-time detection [44]- [90].

### D. Invalidated evaluation methodology

The quantitative test depends on metrics to measure the ability of a system. The conventional metrics are used to estimate the ability of a NIDS to detect attacks as accuracy and detection rate only for the multi-classification tasks. Those metrics do not have the benchmarking criteria for evaluating and non-reflecting the experimental validation of the anomaly-NIDS [86] - [100].

### E. High Computation of DL-NIDS Model

Most troubleshooting systems are complex models that need to optimize time to process data traffic. Few DL-based methodologies are implemented based on GPU technology, while multi-core or Duo core CPUs are more used to implement DL-NIDS methodologies. This accelerated execution the NIDS causes to overload in the CPU (computation time issue) because a limited memory and low speed of CPU for the data processing. This trend is not proportional to the DL and continuously increases in the volume of data.

## VIII. Future Trends

Upcoming research in this field has a lot of potential, especially in anomaly and intrusion detection utilizing DL and ML methods, which are addressed below:

### A. Anomaly-NIDS Effectiveness

As noted, the effectiveness of a NIDS still has false positives, and there can be solutions to reduce false alarms. An unbalanced class problem can be solved with an up-to-date and balanced dataset. To balance the dataset, efficient strategies can boost the number of minority attack cases, or cost-sensitive learning can enhance the training model for unusual attacks [75] [76] . Also, an anomaly-NIDS should have a mechanism to keep the model rules redefined using the dataset for the scenario of attacks to reduce false alarms. Therefore, a few studies are focused on the nature of data and proposed algorithms to implement their methodologies[21] .Most traffic is a sequential data type that needs further attention to suggest a proper algorithm concerning this type, such as an RNN algorithm or a hybrid DL model using feature learning and an ML classifier. These solutions can improve the effectiveness of an anomaly-NIDS with few false alarms.

### B. Light Anomaly-NIDS

In 2025, the number of phones linked to the Internet will exceed 18.25 billion, with a traffic size exceeding 79.4 zettabytes. The architectures of DL insert extra difficulties for a NIDS to be achieved with acceleration in a real-time system. Therefore, high-performance GPUs are used as one solution for quickly and efficiently processing big datasets to handle such difficulty. These GPUs are generally costly. Furthermore, technological and economic constraints impose tight limitations on data's rapidly increasing volume and complexity. Therefore, we have to trade-off the effectiveness and cost, and low-cost GPU platforms could be available to train a DL-NIDS. Another approach to this problem is to use intelligent feature engineering and meta-heuristic techniques to lower the temporal complexity of the DL method. The exemplary architecture of the DL model and some of the hidden layers, their neurons, and optimal weights will yield almost the same great accuracy as if the entire set of features were utilized. Simultaneously, the model’s complexity will be reduced and use fewer computing resources in a real-time setting.

### C. New Threat Detection

According to the latest industrial report, most anomaly-NIDS models were trained and tested based on commonly used datasets and were proposed by different institutions. According to the Panda report, there are new attacks to threaten the network. Most of the studies are not focused on the detection by an ML or DL NIDS model of new threats and are described as follows:

- Ransomware is malware infecting different networks infrastructure parts, such as media outlets, health care, academia, and industry experts. The last report from Apex Laboratory [77] shows ransomware attacks damaged the most significant power utility company, mortgage loan servicing company, and real estate agency in Columbia. It indicates that ransomware attacks injected different domains in network security in many countries. Cybercrime is expected to cost $6 trillion this year (up from $3 trillion in 2015). We anticipate an upsurge in ransomware attacks, with novel versions becoming more disruptive and sophisticated.
- Data-poisoning attack is a novel attack against machine learning techniques by poisoning the training dataset with a few samples of another type of dataset class. It will lead to building a bad AI predictor besides the prediction class as a false diagnosis.
- Zero-day attacks, remote access controls, and insider attacks in cyber-physical attacks target industrial control systems, including industrial factories’ infrastructure. Therefore, ICS might destroy or stop the services of factories, as has happened with water distribution and power stations. Nevertheless, research in this stage is still in its infancy, and additional study is needed to discover and build reliable DL-based NIDS for SCADA networks.

## IX. CONCLUSION

This paper gives a thorough analysis of a network intrusion detection system relying on DL and ML methodologies, intending to allow fresh scholars to update their knowledge regarding prior cyberattacks utilizing a NIDS. The collection of relevant publications in the AI-NIDS study realm is conducted using a systematic methodology. The IDS background theory, as well as its approaches, are first examined. The efficiency of intrusion detection and the complexity of ML/DL techniques are then assessed for each study methodology. According to this study, recent patterns reveal how DL techniques may be utilized to improve the performance of a NIDS in regards to confusion matrix and detection accuracy. Based on the statistical survey, DL techniques were employed in roughly 52% of the suggested solutions, with DNN and AE accounting for 27% of the total, while RNN, DBN, and CNN were employed less often.

Meanwhile, conventional ML is used in 48% of solutions to propose algorithms for an anomaly-NIDS model, with a high of 7% for SVM and less for each DT, RF, and K-mean. A few studies use anomaly-IDS methodologies as hybrid algorithms DL as feature extraction, with ML as anomaly models. Most of the studies that used the DL approach have the topmost effectiveness compared to the ML methods for anomaly-NIDS via the capability to learn essential features by themselves. Furthermore, the study found that 59% of the suggested procedures were verified utilizing out-of-date datasets, which are NSL-KDD and KDD CUP 99. In contrast, in 41% of the techniques, a realistic dataset was utilized for training and testing the anomaly model. A high percentage uses a pretty old dataset. This issue must be addressed to fulfill the realistic data with real-time environments to improve the NIDS performance.

Moreover, there is a lack of studies focused on the complexity of the DL methodology of the NIDS and advanced methods, such as multi-GPU in cloud computing, which are used to reduce training time. These are effective ways to decrease the complexity of anomaly-NIDS models. In the future, we advocate developing a unique, lightweight, and practical DL-NIDS that can identify novel network attacks, for instance, cyber-physical systems, CAN, real-time, and ransomware attacks. Furthermore, we aim to use a statistical measure that quantifies the strength of a relationship or difference between variables in a study. Effect size typically provides a standardized way of assessing the practical or substantive significance of the observed effect. It provides a meaningful and interpretable measure of the size or impact of an effect beyond statistical significance. It helps understanding the practical importance or relevance of the findings, irrespective of sample size or statistical significance.

**APPENDIX A: Summarized information of reviewed articles. The symbol * refers no values/data reported by authors.**

| Ref. | Validation Methodology | | | | | | | | | | | | | | |
|---|---|---|---|---|---|---|---|---|---|---|---|---|---|---|---|
| | Proposed ML/DL for NIDS | | Dataset | Detected Attacks | Effectiveness | | | | | Classification Task | | Hardware Implementation | | Time-based on Methodology | |
| | | | | | Classification Report | | | Accuracy | Detection | Binary | Multiclass | CPU | GPU | Time Training | Time Testing |
| | | | | | Precision | Recall | F-score | | | | | | | | |
| [78] | SVM | DNN | NSL-KDD | 5- Classes | * | * | * | 1.00 | * | × | ✓ | Core i5 | × | × | × |
| [79] | DT | × | KDD CUP 99 | 5- Classes | 0.99 | 0.98 | 0.79 | 1.00 | * | × | ✓ | Core Duo | × | × | × |
| [80] | RF | × | KDD CUP 99 | Normal | 0.98 | * | * | * | * | × | ✓ | Core i5 | × | × | 7.98 |
| | | | | DoS | 0.91 | * | * | * | * | | | | | | |
| | | | | U2R | 0.17 | * | * | * | * | | | | | | |
| | | | | R2L | 0.66 | * | * | * | * | | | | | | |
| | | | | Probe | 0.56 | * | * | * | * | | | | | | |
| [81] | k-means | × | KDD CUP 99 | 5- Classes | * | * | * | * | * | × | ✓ | × | × | × | × |
| [82] | × | LS-SVM | KDD CUP 99 | 2- Classes | * | * | * | 0.96 | 0.95 | ✓ | | × | | × | × |
| [83] | K-means | × | KDD CUP 99 | Normal | 0.96 | 0.85 | 0.92 | * | 0.86 | × | ✓ | × | × | × | × |
| | | | | DoS | 0.97 | 0.97 | 0.97 | * | 1.00 | | | | | | |
| | | | | U2R | 0.75 | 0.73 | 0.75 | * | 0.98 | | | | | | |
| | | | | R2L | 0.65 | 0.92 | 0.73 | * | 0.91 | | | | | | |
| | | | | Probe | 0.85 | 0.95 | 0.92 | * | 0.73 | | | | | | |
| [84] | DT | × | KDD CUP 99 | 22-Classe | * | * | * | 0.20 | * | × | ✓ | Core i7 | × | × | × |
| [85] | NB | × | KDD CUP 99 | Normal | * | * | * | * | 0.98 | × | ✓ | Core i5 | × | 1467 sec | 792 sec |
| | | | | DoS | * | * | * | * | 0.97 | | | | | | |
| | | | | U2R | * | * | * | * | 0.75 | | | | | | |
| | | | | R2L | * | * | * | * | 0.75 | | | | | | |
| | | | | Probe | * | * | * | * | 0.94 | | | | | | |
| [86] | K-Centroid | × | KDD CUP 99 | Normal | * | * | * | 0.86 | * | × | ✓ | × | × | × | × |
| | | | | DoS | * | * | * | 0.92 | * | | | | | | |
| | | | | U2R | * | * | * | 0.96 | * | | | | | | |
| | | | | R2L | * | * | * | 0.77 | * | | | | | | |
| | | | | Probe | * | * | * | 0.32 | * | | | | | | |
| [87] | SVM | × | NSL-KDD | 5- Classes | 0.94 | 0.98 | 0.96 | * | * | × | ✓ | × | × | 2100 sec | 1171 sec |
| [88] | SVM | × | KDD CUP 99 | 5- Classes | * | * | * | 0.90 | * | × | ✓ | × | × | × | × |
| [89] | SVM | × | NSL-KDD | Normal | * | * | * | 0.95 | * | × | ✓ | × | × | × | × |
| | | | | DoS | * | * | * | 0.99 | * | | | | | | |
| | | | | U2R | * | * | * | 0.99 | * | | | | | | |
| | | | | R2L | * | * | * | 0.70 | * | | | | | | |
| | | | | Probe | * | * | * | 1.00 | * | | | | | | |
| [90] | × | AE | NSL-KDD | 5-Classes | 0.86 | 0.96 | 0.76 | * | * | × | ✓ | × | × | × | × |
| [91] | × | SAE | NSL-KDD | 2- Classes | 0.85 | 0.93 | * | 0.87 | * | ✓ | × | × | × | × | × |
| [92] | NN | × | KDD CUP 99 | 2- Classes | * | * | * | 0.85 | * | ✓ | × | core i7 | × | × | × |
| [93] | RepTree | × | UNSW-NB15 | 2- Classes | * | * | * | 0.90 | * | ✓ | × | core i5 | | 2.69 sec | 0.37 sec |
| [94] | EML | × | UNSW-NB15 | 2- Classes | * | * | * | 0.83 | * | ✓ | × | CPU | × | 1.2 min | × |
| [95] | × | GAN | UNSW-NB15 | 10-Classes | * | * | * | 0.99 | * | * | ✓ | Core i5 | × | × | × |
| [96] | LR | × | KDD CUP 99 | 5-Classes | * | * | * | * | 0.99 | * | ✓ | Intel Core Duo | × | × | × |
| [97] | RF | × | CIDDS-001 | 2- Classes | * | * | * | 1.00 | 1.00 | ✓ | * | Core i5 | × | × | × |
| [98] | skip-gram | × | UNSW-NB15 | 2- Classes | 0.99 | 0.82 | * | * | 0.91 | ✓ | * | × | × | × | × |
| [99] | Extra-Tree | * | UNSW-NB15 | Normal | * | * | * | * | 0.94 | ✓ | * | × | × | × | × |
| | | | | DDoS | * | * | * | * | | | | | | | |
| [100] | SVM | * | UNSW-NB15 | 2- Classes | * | * | * | * | 0.96 | ✓ | * | × | × | × | × |
| [101] | Ensemble | * | CIDDS-001 | 2- Classes | * | * | * | * | 0.94 | ✓ | * | × | × | × | × |
| [102] | DT | * | UNSW-NB15 | Normal | 0.93 | 0.93 | * | 0.47 | * | * | ✓ | × | × | × | × |
| | | | | Generic | 0.98 | 0.82 | * | | | | | | | | |
| | | | | Exploits | 76.22 | 0.77 | * | | | | | | | | |
| | | | | Fuzzers | 0.74 | 0.65 | * | | | | | | | | |
| [103] | * | LSTM | UNSW-NB15 | 5-Classes | 0.91 | 0.92 | 0.92 | 0.90 | * | * | ✓ | × | × | × | × |
| [104] | * | GB | UNSW-NB15 | 2- Classes | * | * | * | * | 1.00 | ✓ | * | × | × | × | × |
| [105] | FR | * | UNSW-NB15 | Normal | 0.897 | 0.96 | 0.93 | * | * | * | ✓ | core i5 | × | × | × |
| | | | | Generic | 0.998 | 0.96 | 0.98 | | | | | | | | |
| | | | | Exploits | 0.759 | 0.66 | 0.70 | | | | | | | | |
| | | | | Fuzzers | 0.942 | 0.38 | 0.54 | | | | | | | | |
| | | | | Recon. | 0.888 | 0.82 | 0.85 | | | | | | | | |
| | | | | DoS | 0.351 | 0.46 | 0.39 | | | | | | | | |
| | | | | Analysis | 0.046 | 0.06 | 0.05 | | | | | | | | |
| | | | | Backdoor | 0.151 | 0.40 | 0.21 | | | | | | | | |
| | | | | Shellcode | 0.352 | 0.78 | 0.48 | | | | | | | | |
| | | | | Worms | 0.778 | 0.79 | 0.78 | | | | | | | | |
| [106] | ANN | * | UNSW-NB15 | Normal | 0.989 | 0.99 | 0.99 | * | * | ✓ | × | × | × | × | × |
| | | | | abnormal | | | | | | | | | | | |
| [107] | GMM | * | CICIDS 2017 | BENIGN | * | * | * | 1.00 | * | ✓ | × | × | × | × | × |
| | | | | DDoS | * | * | * | | | | | | | | |
| [108] | K-NN | * | CICIDS 2017 | All types of Attacks | 0.99 | 0.99 | 0.99 | * | 0.99 | * | × | × | × | × | × |
| [109] | * | GAN | CICIDS 2017 | BENIGN | 0.99 | 0.99 | 1.00 | * | * | * | ✓ | × | × | × | × |
| | | | | Brute Force | | | | | | | | | | | |
| | | | | SQL Injection | | | | | | | | | | | |
| | | | | XSS | | | | | | | | | | | |
| | | | | BENIGN | 0.99 | 99.9 | 1.00 | * | * | ✓ | * | | | | |
| | | | | PortScan | | | | | | | | | | | |
| | | | | BENIGN | 0.86 | 0.53 | 65.77 | * | * | ✓ | * | | | | |
| | | | | Bot | | | | | | | | | | | |
| | | | | BENIGN | 100 | 0.60 | 0.75 | * | * | ✓ | * | | | | |
| | | | | Infiltration | | | | | | | | | | | |
| | | | | BENIGN | * | * | * | * | * | * | ✓ | | | | |
| | | | | FTP-Patator | 1.00 | 0.99 | 0.99 | * | * | | | | | | |
| | | | | SSH-Patator | 1.00 | 1.00 | 1.00 | * | * | | | | | | |
| | | | | BENIGN | * | * | * | * | * | * | ✓ | | | | |
| | | | | DoS GoldenEye | 1.00 | 0.99 | 0.99 | * | * | | | | | | |
| | | | | DoS Hulk | 1.00 | 0.99 | 0.99 | * | * | | | | | | |
| | | | | DoS Slowhttptest | 0.99 | 0.99 | 0.99 | * | * | | | | | | |
| | | | | Heartbleed | 1.00 | 1.00 | 1.00 | * | * | | | | | | |
| | | | | Benign | 99.90 | 99.9 | 99.9 | * | * | ✓ | * | | | | |
| | | | | DDoS | | | | | | | | | | | |
| [110] | RF | * | CICIDS 2017 | All types of Attacks | * | * | * | 0.99 | 1.00 | * | ✓ | × | × | × | × |
| [40] | RF | * | CICIDS 2017 | DDoS | * | * | * | * | 0.96 | * | ✓ | Core 2 duo | ✗ | 27.36 sec | ✗ |
| [111] | * | RNN | CICIDS 2017 | * | * | * | * | * | 0.98 | * | ✓ | Intel Xeon | GPUs | 14 min | 1.3 sec |
| [112] | * | RNN | CICIDS 2017 | BENIGN | * | * | * | * | * | * | * | CPU | GPU | 20.7 sec | 7.21 sec |
| | | | | Brute Force | * | * | * | * | 0.82 | | | | | | |
| | | | | SQL Injection | * | * | * | * | 0.82 | | | | | | |
| | | | | XSS | * | * | * | * | 0.92 | | | | | | |
| | | | | BENIGN | * | * | * | * | * | ✓ | * | | | | |
| | | | | PortScan | * | * | * | * | 0.90 | | | | | | |
| | | | | BENIGN | * | * | * | * | * | ✓ | * | | | | |
| | | | | BoT | * | * | * | * | 0.98 | | | | | | |
| | | | | BENIGN | * | * | * | * | * | ✓ | * | | | | |
| | | | | Infiltration | * | * | * | * | 1.00 | | | | | | |
| | | | | BENIGN | * | * | * | * | * | * | ✓ | | | | |
| | | | | FTP-Pastor | * | * | * | * | 1.00 | | | | | | |
| | | | | SSH-Pastor | * | * | * | * | 1.00 | | | | | | |
| | | | | BENIGN | * | * | * | * | 0.98 | * | ✓ | | | | |
| | | | | DoS GoldenEye | * | * | * | * | 0.77 | | | | | | |
| | | | | DoS Hulk | * | * | * | * | 0.98 | | | | | | |
| | | | | DoS Slowhttptest | * | * | * | * | 0.94 | | | | | | |
| | | | | Heartbleed | * | * | * | * | 1.00 | | | | | | |
| | | | | BENIGN | * | * | * | * | 1.00 | ✓ | * | | | | |
| | | | | DDoS | * | * | * | * | | | | | | | |
| [113] | * | RF | CICIDS 2017 | Normal | 1.00 | 1.00 | 1.00 | 1.00 | | ✗ | ✓ | core i3 | GPU | 9 sec | 8 sec |
| | | | | Brute-Force | 1.00 | 0.78 | 0.77 | 0.76 | | | | | | | |
| | | | | XSS | 0.56 | 0.47 | 0.49 | 0.49 | | | | | | | |
| | | | | SQL-injection | 0.22 | 1.00 | 0.22 | 0.36 | | | | | | | |
| [114] | * | AE | KDD CUP 99 | Normal, Abnormal | 0.99 | ✗ | ✗ | ✗ | ✗ | ✓ | ✗ | ✗ | ✗ | ✗ | ✗ |
| [115] | * | CNN, AE | NSL-KDD | Normal | 1.00 | ✗ | ✗ | ✗ | ✗ | ✗ | ✓ | core i5 | ✗ | ✗ | ✗ |
| | | | | DoS | 1.00 | ✗ | ✗ | ✗ | ✗ | | | | | | |
| | | | | U2R | 0.70 | ✗ | ✗ | ✗ | ✗ | | | | | | |
| | | | | R2L | 0.70 | ✗ | ✗ | ✗ | ✗ | | | | | | |
| | | | | Probe | 1.00 | ✗ | ✗ | ✗ | ✗ | | | | | | |
| [116] | * | Bi-RNN | UNSW-NB15 | Normal, Attack | 0.95 | 1.00 | ✗ | ✗ | ✗ | ✓ | ✗ | Core i7 | * | 3228 sec | 54 sec |
| [117] | * | DNN | Realistic data | Normal, Abnormal | 0.99 | ✗ | ✗ | ✗ | ✗ | ✓ | ✗ | Core i7 | ✗ | ✗ | ✗ |
| [118] | * | AE | NSL-KDD | Normal | 0.97 | 0.70 | 0.97 | 0.81 | ✗ | ✗ | ✓ | Intel Xeon | ✗ | ✗ | ✗ |
| | | | | DoS | 0.83 | 0.96 | 0.83 | 0.89 | ✗ | | | | | | |
| | | | | U2R | 0.11 | 0.81 | 0.32 | 0.64 | ✗ | | | | | | |
| | | | | R2L | 0.32 | 0.73 | 0.11 | 0.18 | ✗ | | | | | | |
| | | | | Probe | 0.80 | 0.86 | 0.80 | 0.46 | ✗ | | | | | | |
| [119] | * | SAE | KDD CUP 99 | Normal, Attack | 0.95 | ✗ | ✗ | ✗ | ✗ | ✓ | ✗ | ✗ | GPU | ✗ | ✗ |
| [120] | * | SAE | NSL-KDD | Normal | 0.97 | 1.00 | 0.97 | 0.98 | ✗ | ✗ | ✓ | ✗ | GPU | ✗ | ✗ |
| | | | | DoS | 0.94 | 1.00 | 0.94 | 0.97 | ✗ | | | | | | |
| | | | | U2R | 0.94 | 1.00 | 0.94 | 0.97 | ✗ | | | | | | |
| | | | | R2L | 0.03 | 1.00 | 0.03 | 0.07 | ✗ | | | | | | |
| | | | | Probe | 0.94 | 1.00 | 0.72 | 0.97 | ✗ | | | | | | |
| [121] | * | DNN | NSL-KDD | Normal, DoS, U2R, R2L, Probe | 0.86 | ✗ | ✗ | ✗ | ✗ | ✗ | ✓ | Core i3 | ✗ | ✗ | ✗ |
| [122] | * | DNN | AWID | normal, flooding, impersonation, and injection | 0.99 | ✗ | ✗ | ✗ | ✗ | ✗ | ✓ | Core i3 | ✗ | ✗ | ✗ |
| [123] | DNN | AE | KDD CUP 99 | Normal, DoS,U2R, R2L, Probe | 0.99 | 0.99 | 0.99 | ✗ | ✗ | ✗ | ✓ | core i3 | ✗ | ✗ | ✗ |
| [124] | * | DNN | UNSW-NB15 | Normal, Attack | 0.92 | ✗ | ✗ | ✗ | ✗ | ✗ | ✓ | Core i5 | ✗ | ✗ | ✗ |
| [125] | * | DNN | CICIDS 2017 | Normal, Attack | 0.99 | 0.99 | 0.98 | ✗ | ✗ | ✓ | ✗ | Core i3 | ✗ | ✗ | ✗ |
| [126] | * | WDL STM, CNN | UNSW-NB15 | Normal | 0.98 | 1.00 | 1.00 | 1.00 | * | ✗ | ✓ | Core i7 | ✗ | ✗ | ✗ |
| | | | | DoS | | 0.64 | 0.80 | 0.71 | * | | | | | | |
| | | | | Exploits | | 0.32 | 0.27 | 0.29 | * | | | | | | |
| | | | | Backdoor | | 0.50 | 0.07 | 0.12 | * | | | | | | |
| | | | | Analysis | | 0.44 | 0.09 | 0.15 | * | | | | | | |
| | | | | Fuzzers | | 0.71 | 0.61 | 0.66 | * | | | | | | |
| | | | | Generic | | 1.00 | 0.99 | 0.99 | * | | | | | | |
| | | | | Reconnaissance | | 0.93 | 0.77 | 0.84 | * | | | | | | |
| | | | | Shellcode | | 0.82 | 0.79 | 0.81 | * | | | | | | |
| | | | | Worms | | 0.50 | 0.09 | 0.15 | * | | | | | | |
| [127] | SVM | AE | ISCX 2012 | Normal, Attack | 0.994 | 0.99 | 0.994 | 0.99 | × | ✓ | × | Core i7 quad-core | × | × | × |
| [128] | * | GAN | NSL-KDD | Normal, Attack | 0.89 | 0.89 | 0.90 | 0.10 | × | ✓ | × | × | GPU | × | × |
| [129] | * | CNN-RNN-LSTM | CIDDS-001 | Normal, malicious | 0.91 | 0.98 | 0.68 | 0.80 | × | ✓ | × | Core i3 | GPU | × | × |
| [130] | * | DAE-MLP | UNSW-NB15 | Normal | * | 0.99 | 0.99 | 0.99 | * | ✓ | × | × | × | × | × |
| | | | | Attack | * | 0.95 | 0.94 | 0.95 | * | | | | | | |
| [131] | * | CNN | NSL-KDD | Normal, DoS U2R, R2L, Probe | × | × | × | × | × | × | ✓ | × | × | × | × |
| [132] | * | RNN | UNSW NB15 | DDoS DoS Reconnaissance Normal Theft | × | 0.93 | × | × | × | × | ✓ | × | × | × | × |
| [133] | * | GAN | AWID | flooding | 0.993 | 0.74 | 0.921 | 0.618 | × | × | ✓ | × | × | × | × |
| | | | | impersonation | 0.965 | 0.00 | 0.00 | 0.00 | | | | | | | |
| | | | | injection | 0.998 | 0.96 | 0.934 | 0.999 | | | | | | | |
| | | | | normal | 0.957 | 0.97 | 0.95 | 0.99 | | | | | | | |
| [134] | * | LSTM | Real Time | Normal, Abnormal | × | × | × | × | | ✓ | × | × | GPU | × | × |
| [135] | * | DNN | CIDS | Normal, malicious | 0.96 | 0.91 | 0.91 | 0.91 | × | ✓ | × | core i5 | × | × | × |
| [136] | SVM | DBN | NSL-KDD | Normal, Attack | 0.974 | 0.97 | 0.977 | 0.97 | * | ✓ | * | core i7 | * | * | * |
| [137] | RF | * | CIDDS-001 | Normal, malicious | 0.99 | × | × | × | × | × | × | × | GPU | × | × |
| [55] | * | DNN | CICIDS 2017 | Normal | 0.56 | × | × | × | × | × | ✓ | × | GPU | × | × |
| | | | | SSH-Patater | 0.95 | × | × | × | × | | | | | | |
| | | | | FTP-Patater | 0.92 | × | × | × | × | | | | | | |
| | | | | Web | 0.98 | × | × | × | × | | | | | | |
| | | | | Bot | 0.95 | × | × | × | × | | | | | | |
| | | | | DDoS | 0.85 | × | × | × | × | | | | | | |
| | | | | Portscan | 0.85 | × | × | × | × | | | | | | |
| [138] | * | DNN | Realistic data | Blackhole | × | 0.97 | | 0.97 | × | × | ✓ | × | × | × | × |
| | | | | Opportunistic Service | × | 0.95 | 0.98 | 0.97 | × | | | | | | |
| | | | | DDoS | × | 0.96 | 0.98 | 0.96 | × | | | | | | |
| | | | | Sinkhole | × | 0.99 | 0.99 | 0.99 | × | | | | | | |
| | | | | Wormhole | × | 0.96 | 0.97 | 0.98 | × | | | | | | |
| [139] | * | RBM | Realistic data | Normal, malicious | 0.99 | × | × | × | × | ✓ | × | × | × | × | × |
| [140] | k-mean | RBM | NSL-KDD | Normal, malicious | 0.92 | × | × | × | × | × | × | Core Due | × | × | × |
| [141] | RF | * | NSL-KDD | Normal | 1.00 | × | × | × | × | × | ✓ | Core i7 | GPU | × | × |
| | | | | DDOS | 1.00 | × | × | × | | | | | | | |
| | | | | DOS | 0.98 | × | × | × | | | | | | | |
| | | | | Reconnaissance | 1.00 | × | × | × | | | | | | | |
| | | | | Theft | 0.96 | × | × | × | | | | | | | |
| [142] | × | DBN, PNN | KDD CUP 99 | Normal, Abnormal | 0.99 | × | × | × | × | ✓ | × | × | GPU | × | × |
| [29] | × | RBM | ISCX 2012 | Normal, Abnormal | 0.88 | × | × | × | × | ✓ | × | × | × | × | × |
| [143] | × | RBM | KDD CUP 99 | Normal, DDoS | 0.99 | × | × | 0.99 | × | ✓ | × | × | × | × | × |
| [144] | × | AE | NSL-KDD | Normal | × | 0.85 | 0.96 | 0.90 | × | × | ✓ | Core i7 | GPU | × | × |
| | | | | DoS | × | 0.95 | 0.98 | 0.97 | × | | | | | | |
| | | | | Probe | × | 0.69 | 0.94 | 0.80 | × | | | | | | |
| | | | | R2L | × | 0.99 | 0.39 | 0.56 | × | | | | | | |
| | | | | U2R | × | × | × | × | × | | | | | | |
| [145] | RF | × | CIC-IDS2017 | BENIGN | 1.00 | 1.00 | 1.00 | × | × | * | ✓ | Core i5 | 8 GB | 89.9 | 1.22 |
| | | | | DoS GoldenEye | 1.00 | 1.00 | 1.00 | × | × | | | | | | |
| | | | | DoS Hulk | 1.00 | 1.00 | 1.00 | × | × | | | | | | |
| | | | | DoS Slowhttptest | 1.00 | 1.00 | 1.00 | × | × | | | | | | |
| | | | | Heartbleed | 1.00 | 1.00 | 1.00 | × | × | | | | | | |
| [146] | SVM. | × | UNSW-NB15 | Normal | × | × | 0.60 | × | × | * | ✓ | Core(TM) i5 | 4GB | × | × |
| | | | | DoS | × | × | 0.0 | × | × | | | | | | |
| | | | | Exploits | × | × | 0.80 | × | × | | | | | | |
| | | | | Backdoor | × | × | 0.00 | × | × | | | | | | |
| | | | | Analysis | × | × | 0.00 | × | × | | | | | | |
| | | | | Fuzzers | × | × | 0.64 | × | × | | | | | | |
| | | | | Generic | × | × | 0.96 | × | × | | | | | | |
| | | | | Reconnai | × | × | 0.36 | × | × | | | | | | |
| | | | | Shellcode | × | × | 0.81 | × | × | | | | | | |
| | | | | Worms | × | × | 0.00 | × | × | | | | | | |
| [147] | × | AE | NSL-KDD | All classes | × | × | × | 0.93 | 0.99 | × | ✓ | Core i7 | 8GB | × | × |
| [148] | × | AE | privet | Normal | × | × | × | × | × | ✓ | × | × | × | × | × |
| | | | | Abnormal | | | | | | | | | | | |
| [149] | × | AE | KDD CUP'99 | All classes | 1.00 | 1.00 | 1.00 | 1.00 | × | × | ✓ | Core i7 | 32 GB | × | × |
| [150] | × | RNN | CICIDS -2017. | All classes | 0.99 | 0.99 | 0.99 | 0.99 | 0.97 | × | ✓ | × | × | × | × |
| [151] | × | DBN | KDD CUP 99 | All classes | × | × | × | 0.90 | × | × | ✓ | Core i7 | 16 GB | | 51.07 |
| [152] | × | AE | CICIDS 2018 | Benign | 99.7 | 99.9 | 99.8 | × | × | × | ✓ | Core i9 | 128 GB | × | 3885 |
| | | | | Bot | 1.00 | 0.99 | 1.00 | × | × | | | | | | |
| | | | | Web | 0.85 | 0.94 | 0.90 | × | × | | | | | | |
| | | | | XSS | 1.00 | 0.74 | 0.85 | × | × | | | | | | |
| | | | | HOIC | 1.00 | 1.00 | 1.00 | × | × | | | | | | |
| | | | | LOIC-UDP | 0.96 | 0.99 | 0.98 | × | × | | | | | | |
| | | | | LOIC-UDP | 0.96 | 0.99 | 0.98 | × | × | | | | | | |
| | | | | LOIC-HTTP | 1.00 | 1.00 | 1.00 | × | × | | | | | | |
| | | | | GoldenEye | 1.00 | 1.00 | 1.00 | × | × | | | | | | |
| | | | | Hulk | 1.00 | 1.00 | 1.00 | × | × | | | | | | |
| | | | | SlowHTTPTest | 1.00 | 0.97 | 98.4 | × | × | | | | | | |
| | | | | Slowloris | 1.00 | 99.9 | 1.00 | × | × | | | | | | |
| | | | | BruteForce | 98.4 | 100 | 1.00 | × | × | | | | | | |
| | | | | Bot-Iot | 99.1 | 0.98 | 0.98 | × | × | | | | | | |
| | | | | SQL Injection | 1.00 | 0.60 | 0.74 | × | × | | | | | | |
| | | | | Bruteforce | 1.00 | 1.00 | 1.00 | × | × | | | | | | |
| [153] | * | PNN | × | All classes | 0.99 | | 0.99 | 0.99 | × | × | ✓ | × | × | × | × |
| [154] | RF | × | UNSW-NB15 | All classes | × | × | × | 0.83 | × | × | ✓ | × | × | × | × |
| [155] | xgboost | × | CICIDS 2017 | All classes | × | × | × | 0.99 | × | × | ✓ | Core i5 | × | 8GB | × |
| [156] | EM | × | CSE-CIC-IDS2018 | All classes | × | × | × | 0.900 | × | × | ✓ | × | × | × | × |
| [157] | × | AE | BoT-IoT | Normal | × | × | × | 1.00 | × | × | ✓ | × | × | 2012.6 | × |
| | | | | Information Gathering | × | × | × | 1.00 | × | | | | | | |
| | | | | DoS | × | × | × | 1.00 | × | | | | | | |
| | | | | DDoS | × | × | × | 1.00 | × | | | | | | |
| | | | | Information Theft | × | × | × | 1.00 | × | | | | | | |
| [158] | AdaBoost | × | CICIDS 2017 | Binary-class | × | × | × | × | × | ✓ | × | × | × | × | 131,9 |
| [159] | RF | × | CIC-IDS2017 | Multi-class | 0.99 | 0.94 | 0.96 | 0.99 | × | × | ✓ | Intel Pentium, CPU B960 | 4 GB | × | 455.317 |
| [160] | × | CNN | CIC-IDS2017 | Binary-class | 0.99 | 0.99 | 0.99 | 0.99 | × | ✓ | × | Core i5 | 8 GB | 39.52 | 0.061 |
| [161] | × | AE | CIC-IDS2017 | Multi-class | 0.99 | 0.99 | 0.99 | 0.99 | × | × | ✓ | × | × | × | 0.09 |
| [162] | × | LSTM | CIC-IDS2017 | Multi-class | 1.00 | 1.00 | 1.00 | 1.00 | × | × | ✓ | × | × | 9.20 | × |
| [163]v | × | CNN | CIC-IDS2017 | Multi-class | × | × | × | × | 0.90 | × | ✓ | Core i7 | 16 GB | 120 | 180 |
| [164]v | × | CNN | CIC-IDS2017 | Multi-class | × | × | × | 0.99 | × | × | ✓ | × | × | × | × |
| [165] | × | AE | CIC-IDS2017 | Normal | 0.99 | 0.99 | 0.97 | × | × | × | ✓ | Core i7 | 16 GB | 1851.02 | 0.43 |
| | | | | SSH | 0.99 | 0.50 | 0.67 | × | × | | | | | | |
| | | | | FTP-Patater | 0.98 | 0.93 | 0.96 | × | × | | | | | | |
| | | | | Web | 0.98 | × | × | × | × | | | | | | |
| | | | | DoS GoldenEye | 0.99 | 0.88 | 0.93 | × | × | | | | | | |
| | | | | DoS Hulk | 0.72 | 0.73 | 0.81 | | | | | | | | |
| | | | | Bot | 0.0 | 0.00 | 0.00 | × | × | | | | | | |
| | | | | DDoS | 0.99 | 0.92 | 0.95 | × | × | | | | | | |
| | | | | Portscan | 0.85 | × | × | × | × | | | | | | |
| | | | | infiltration | 0.00 | 0.00 | 0.00 | × | × | | | | | | |